\documentclass{aa}  

\usepackage{graphicx}
\usepackage{orcidlink}
\usepackage{txfonts}
\usepackage{lipsum}
\usepackage{subcaption}         
\usepackage{lscape}             
\usepackage{placeins}           

\usepackage{booktabs}
\usepackage[]{hyperref}

\begin{document}

   \title{The quiescent states of V745~Sco and V3890~Sgr:\\VLT/X-shooter and Swift/XRT+UVOT observations}
   \titlerunning{Quiescent states of V745~Sco \& V3890~Sgr}

   \author{Bor Jamnik\orcidlink{0009-0008-9261-0418}
          \inst{1}\corrauth{bor.jamnik@fmf.uni-lj.si},
          Ulisse Munari\orcidlink{0000-0001-6805-9664}\inst{2}\email{ulisse.munari@inaf.it},
          Nicola Masetti\orcidlink{0000-0001-9487-7740}\inst{3,4}\email{nicola.masetti@inaf.it},
          Gregor Traven\orcidlink{0000-0002-1391-9097}\inst{1}\email{gregor.traven@fmf.uni-lj.si},
          Manica Perko\orcidlink{0000-0002-7043-8038}\inst{1}\email{manica.perko@fmf.uni-lj.si}
          }
   \authorrunning{B. Jamnik et al.}

   \institute{Faculty of Mathematics and Physics, University of Ljubljana, Jadranska 19, 1000 Ljubljana, Slovenia
        \and
        INAF National Institute of Astrophysics, Astronomical Observatory of Padova, 36012 Asiago (VI), Italy
        \and 
        INAF – Osservatorio di Astrofisica e Scienza dello Spazio, via Piero Gobetti 101, I-40129 Bologna, Italy
        \and 
        Instituto de Astrofisica, Depto. de Ciencias Fisicas, Facultad de Ciencias Exactas, Universidad Andres Bello, Av. Fernandez Concha 700, Las Condes, Santiago, Chile
            }
   \date{Received YYY; accepted ZZZ}

 
  \abstract
    {Symbiotic recurrent novae are symbiotic binaries in which multiple thermonuclear outbursts have been observed. Only four such systems are currently known in the Galaxy: T~CrB, RS~Oph, V745~Sco, and V3890~Sgr. They provide convenient laboratories for studying accretion and binary interaction in symbiotic novae during quiescence. We characterise the quiescent states of the less-studied systems V745~Sco and V3890~Sgr and compare their properties with those of T~CrB and RS~Oph. We analyse VLT/X-shooter spectra together with Swift/XRT and UVOT observations, supplemented by ground-based optical photometry and archival data. We determine the spectral types of the red-giant donors and compare the spectral energy distributions of the systems with those of single bulge giants of the same spectral type to investigate the properties of the accretion. We classify the donor stars in V745~Sco and V3890~Sgr as M7III and M6.5III, respectively. We derive reddenings and distances by comparing the systems' colours and magnitudes with those of typical giants in the Galactic bulge. Both objects exhibit a significant ultraviolet excess relative to normal late-type giants, indicating ongoing accretion during quiescence. From this excess we calculate lower limits on the mass transfer rate in the systems to be $\dot{M}_\mathrm{acc}\gtrsim2.4\times10^{-8}\ M_\odot\ \mathrm{yr}^{-1}$ and $\dot{M}_\mathrm{acc}\gtrsim2.5\times10^{-8}\ M_\odot\ \mathrm{yr}^{-1}$ for V745~Sco and V3890~Sgr, respectively. We discuss several evolutionary scenarios for the quiescent state in the systems, which are consistent with such high lower limits of the mass transfer rate.}
  


   \maketitle
\nolinenumbers
%

\section{Introduction}
Symbiotic stars (SySts) are evolved close binary systems, consisting of a compact object (usually a white dwarf, WD, sometimes a neutron star) and a red giant (RG), in which the effect of accretion of mass from the RG onto the compact object is detectable at some wavelength \citep{2013A&A...559A...6L}.

During the lifetime of a SySt, two phases are observed: i) accreting-only phase, spanning several hundreds or thousands of years, during which the mass is accreted onto the WD, and ii) the burning phase, which follows after enough mass for ignition has accumulated on the WD. Burning may occur in either degenerate or non-degenerate conditions, depending on the state of the matter on the WD surface. If the matter is non-degenerate, it burns stably over the course of a few centuries, while ignition of degenerate matter leads to a classical nova outburst. Once the burning depletes hydrogen in the accumulated shell, accretion resumes to prepare for a new cycle. The concept of alternating accreting-only and burning phases, and of the cycle as a whole, was first introduced by \citet{2019arXiv190901389M}. If the accreted matter causes the WD to exceed the Chandrasekhar mass, a SySt can meet the end of its life as a Type Ia supernova.

The list of symbiotic novae (SySts in which a classical nova outburst has been detected) has recently been expanded to 11 objects \citep{2025CoSka..55c..47M}. Four of those are recurrent (symbiotic recurrent novae, SyRNe), meaning that multiple nova outbursts have been detected in the same system on a scale of a human lifetime. These systems are: T~CrB, RS~Oph, \object{V745~Sco} and \object{V3890~Sgr}\footnote{V407~Cyg is also mentioned in the literature as being a candidate SyRN, with some evidence for a 1936 outburst in addition to the recent 2010 one.}. Their short inter-outburst periods (on the order of a few tens of years) provide us with a unique opportunity to observe pre- and post-outburst states of symbiotic systems, which allows us to study the processes leading up to the eruption, the return of the system to quiescence, and the influence of the outburst on the systems during quiescence. The four SyRNe all harbour a near-Chandrasekhar mass WD and an M-type giant, orbiting with periods of a few hundred days.  

All four known SyRNe have been intensely studied from X-ray to radio wavelengths during their outbursts and the consequent returns to quiescence. The quiescent phase of T~CrB has also been monitored since its first detected outburst in 1866 \citep{2023MNRAS.524.3146S}, and the efforts to understand the quiescent phase of the system have increased since, especially after an increase in activity in 2015 hinted at an outburst around 2026 \citep[e.g.][among many others]{munari20162015, 2020ApJ...902L..14L, 2023A&A...680L..18Z, 2025A&A...694A..85P, 2025MNRAS.541L..14M}. The quiescent phase of RS~Oph has also been investigated previously \citep[e.g.][]{2018MNRAS.480.1363Z,2020BlgAJ..33....3G,2025MNRAS.537.2046H, 2026MNRAS.547ag466W}, which can be attributed to the rather short time between outbursts in the system (9--21 years). 

V745~Sco and V3890~Sgr are much fainter than the other two SyRNe, have intermediate inter-outburst periods of about 30--40 years (compared to 9--21 years for RS~Oph and $\sim80$ years for T~CrB), and are so far south that they are out of reach of many northern facilities. All of these factors have contributed to these systems receiving less attention than their brighter counterparts. 

The quiescent state of V745~Sco remains poorly characterised. Although it has been discussed in a couple of broader studies of SyRNe \citep{1999A&A...344..177A, 2001ApJ...558..323H} and in some investigations of its outbursts, the lack of observational data has limited the characterisation of the system to a binary containing a late-type M-giant and a massive white dwarf, which is accreting at a significant level. The quiescent phase of V3890~Sgr is known better. \citet{1999A&A...344..177A} reported a varying level of activity in the system, confirmed by \citet{2021MNRAS.504.2122M}, who studied its behaviour before the 2019 outburst. In this 2021 paper they also reported an accurate orbital solution, based on variation of the system's radial velocity. Another recent paper by \citet{2022MNRAS.517.6064K} focused on characterisation of the giant in the system. 

In this paper we aim to characterise the quiescent states of V745~Sco and V3890~Sgr in 2024. We present our observations in Sect.~\ref{sec:observations} and determine spectral types of the donors, as well as extinction and distances to the systems in Sect.~\ref{sec:sptype_ext}. In Sect.~\ref{sec:SED} we discuss the SEDs of the systems and calculate the mass transfer rates in Sect.~\ref{sec:mass_transfer_rate}. We report prominent spectral features in Sect.~\ref{sec:spectral_features} and conclude in Sect.~\ref{sec:conclusion}.

\section{Observations}\label{sec:observations}
\subsection{X-shooter spectra}
The X-shooter \citep{2011A&A...536A.105V} spectra were acquired on 2024~May~2 for V745~Sco and on 2024~July~31 for V3890~Sgr, as a result of proposal \texttt{113.26NM.001} (PI M. Perko) from the ESO Call for Proposals Period 113.
\par
Both objects were observed in A-B nodding mode. V745~Sco was observed in total for $3660/3472/2400\,$s in UltraViolet-Blue/Visual/Near-InfraRed (UVB/VIS/NIR) arms to achieve S/N of $41.7/168.9/524.5$. V3890~Sgr was observed in total for $848/660/960\,$s for a S/N of $33.4/99.1/514.4$. For both stars slits of 0.5\arcsec/0.4\arcsec/0.4\arcsec\ were used to obtain median resolving powers of $R=\lambda/\delta\lambda=9861/18340/11424$ in the respective arms, which cover 3000--5595/5595--10240/10240--24800 $\AA$ spectral ranges.
\par 
Standard spectrophotometric calibrations that are carried out are only sufficient for relative fluxing of the spectra and for correcting the instrumental response. We did not request simultaneous observations of spectrophotometric standard stars for absolute spectral fluxing. Instead, we use temporally adjacent photometric observations to obtain accurate absolute fluxes of the X-shooter spectra. We compute synthetic magnitudes from the X-shooter spectra and then scale the spectra in such a way that the calculated magnitudes match the observed ones.
\par 
For V3890~Sgr we were able to arrange for photometric observations by the ANS Collaboration to be carried out in the days coinciding with the X-shooter observation to obtain $B$, $V$, $R$ and $I$ band photometry (see Sect.~\ref{sec:BVRI}).
\par 
V745~Sco is unfortunately too far south to be observed by ANS Collaboration. We therefore use AAVSO photometric data that was gathered between 2024~July~30 and 2024~November~5. We observe no statistically significant variability or trends in $V$, $R$ and $I$ bands and thus take the mean value of the magnitudes measured in this period to use in fluxing. In $B$ band we observe a slight increase with time so we fit a linear function and extrapolate to 2024~May~2.
\subsection{Swift observation}
Two target-of-opportunity pointings were performed with the {\it Swift} satellite \citep{2004ApJ...611.1005G} on 2024~May~23 and 2024~August~28 on V745~Sco and V3890~Sgr, respectively, soon after the corresponding VLT/X-shooter observation presented in this paper.

The {\it Swift} data were acquired with the on-board instruments X-Ray Telescope \citep[XRT;][]{2005SSRv..120..165B} and the UltraViolet Optical Telescope \citep[UVOT;][]{2005SSRv..120...95R}. The XRT covers the 0.3--10 keV X-ray band, whereas UVOT data were collected using the $UVW1$ and $UVW2$ ultraviolet filters, with reference wavelengths 2600 and 1928 \AA, respectively \citep[see][for details]{2008MNRAS.383..627P,2011AIPC.1358..373B}. The two instruments observed each source simultaneously; details on each pointing are reported in Table~\ref{tab:SwiftUVOT}. All data were reduced within the {\sc ftools} environment \citep{1995ASPC...77..367B}. Both sources exhibit strong accretion signatures, which is easily seen in the X-shooter spectra (see Fig.~\ref{fig:spectral_fit_V745Sco} and Sect.~\ref{sec:spectral_features}). We therefore expect significant contribution of the accretion in the UV and do not expect any relevant contamination due to the red leak phenomenon that affects the $UVW1$ and $UVW2$ filters in case of observations of very red objects \citep[for details see e.g.][]{2010ApJ...721.1608B,2014AJ....148..131S}.
\begin{table*}
\caption[]{UVOT observation information.}\label{tab:SwiftUVOT}

\begin{center}
\begin{tabular}{lccccc}
\noalign{\smallskip}
\hline
\hline
\noalign{\smallskip}

          & Observation &  \\
Source 	  & start time  &  filter & Exposure & observed [unabsorbed] &  observed [unabsorbed] \\
          &    (UT)     &        & time (s) & magnitude             & flux density ($\times$10$^{-15}$ erg cm$^{-2}$ s$^{-1}$ \AA$^{-1}$) \\

\noalign{\smallskip}
\hline
\noalign{\smallskip}
 
V745~Sco  & 12:38 & $UVW1$ & 670 & $18.68\pm0.14$ [$12.88^{+1.39}_{-1.38}$] & $0.135\pm0.018$ [$28^{+71}_{-20}$] \\
\noalign{\smallskip}
          &       & $UVW2$ & 703 & $19.8\pm0.3$   [$12.7^{+1.7}_{-1.8}$]   & $0.064\pm0.019$ [$46^{+177}_{-38}$] \\

\noalign{\medskip}

V3890~Sgr & 14:58 &  $UVW1$ & 790 & $16.06\pm0.04$ [$12.12\pm0.04$] & $1.5\pm0.06$  [$56\pm3$] \\
\noalign{\smallskip}
          &       & $UVW2$ & 810 & $16.97\pm0.06$ [$12.10\pm0.06$] & $0.88\pm0.04$ [$78\pm4$] \\

\hline
\noalign{\smallskip}
\end{tabular}
\end{center}
\end{table*}

\begin{table}
\caption{XRT flux upper limits.}              
\label{table:SwiftXRT}      
\centering                                      
\begin{tabular}{l c c}          
\noalign{\smallskip}
\hline
\hline
\noalign{\smallskip}

Source 	   & Exposure &         observed [unabsorbed] flux          \\
           & time (s) & ($\times$10$^{-13}$ erg cm$^{-2}$ s$^{-1}$)  \\

\noalign{\smallskip}
\hline
\noalign{\smallskip}
 
V745~Sco  & 1522 & $<$2.3 [$<$2.9]\\
\noalign{\medskip}
V3890~Sgr & 1635 & $<$2.0 [$<$2.5]\\

\hline                      
\end{tabular}
\end{table}
\par
Count rates on Level 2 (i.e., calibrated and containing astrometric information) UVOT images were extracted through aperture photometry within a 5$''$ radius centered on the position of the objects of interest, whereas the corresponding background was evaluated for each image using a combination of several circular regions in source-free nearby areas. The data were then calibrated using the UVOT photometric system described by \citet{2008MNRAS.383..627P}; the most recent fixings (November 2020) recommended by the UVOT team were taken into account and a check to reject small-scale sensitivity inhomogeneities\footnote{\url{ https://swift.gsfc.nasa.gov/analysis/uvot_digest/sss_check.html}} was also performed.
\par 
Both sources were detected with UVOT; their Vega system magnitudes are reported in Table~\ref{tab:SwiftUVOT} together with the corresponding flux densities. The respective values corrected for the colour excess $E(B-V)$ toward each source ($0.87\pm0.20$~mag for V745~Sco and 0.59~mag for V3890~Sgr; see Sect.~\ref{sec:extincition}) following the prescription of \citet[][their Table~5]{2008ApJ...672..787K} are also reported in the table. As the uncertainty of the adopted V3890~Sgr $E(B-V)$ value is not reported in the original work \citep{2019ATel13069....1M}, it is not accounted for in the uncertainties of the dereddened fluxes and magnitudes of V3890~Sgr. The reported formal uncertainties are therefore correspondingly smaller than those of V745~Sco. 
\par 
The XRT data analysis was performed using the {\sc xrtdas} standard pipeline package ({\sc xrtpipeline} v. 0.13.4) in order to produce screened event files.  All X-ray data were acquired in photon counting (PC) mode \citep{2004SPIE.5165..217H} adopting the standard grade filtering (0--12 for PC) according to the XRT nomenclature. Scientific data for each source were extracted from the images using a radius of 47$''$ (20 pixels) centered at the optical coordinates of the source, while the corresponding background was evaluated in a source-free region of radius 94$''$ (40 pixels) within the same XRT acquisition. No emission was detected in the 0.3--10 keV range from any of the two sources: using the {\sc xspec} package and the procedure of \citet{1986ApJ...303..336G}, we determined 3$\sigma$ limit count rates of $4.3\times10^{-3}$ and $4.0\times10^{-3}$~counts~s$^{-1}$ for V745~Sco and V3890~Sgr, respectively.
\par
We then evaluated the corresponding X-ray flux using the {\sc webpimms} online tool\footnote{\url{ https://heasarc.gsfc.nasa.gov/cgi-bin/Tools/w3pimms/w3pimms.pl}} by assuming an {\sc apec} spectral shape \citep{2001ApJ...556L..91S} with Solar metallicity and temperature $7.7\ \mathrm{keV}$, which seems reasonable based on observed X-ray spectra of other symbiotic stars \citep{2013A&A...559A...6L, 2018ApJ...864...46L, 2024MNRAS.532.1421T}, and assuming an intervening hydrogen column density absorptions ($N_{\rm H}$) of $4.8\times10^{21}$ cm$^{-2}$ for V745~Sco and $3.3\times10^{21}$ cm$^{-2}$ for V3890~Sgr (both obtained by again adopting the colour excesses reported in Sect.~\ref{sec:extincition}, combined with the empirical formula of \citealp{1995A&A...293..889P}). The corresponding observed and unabsorbed flux upper limits are reported in Table~\ref{table:SwiftXRT} and are model dependent. Any selection of reasonable temperature of an {\sc apec}, bremsstrahlung, or blackbody model produces an upper limit that differs from the reported one by no more than a factor of 3. Uncertainty of $E(B-V)$ for the case of V745~Sco plays a lesser role, changing observed and unabsorbed flux upper limit by up to 20\%.

\subsection{BVRI photometry}\label{sec:BVRI}
On our request, $BVRI$ photometry of V3890~Sgr has been collected by ANS Collaboration observers around the epoch of the X-shooter observation to help absolute fluxing as provided by the default ESO pipeline. All bands have been observed on each visit to the object and all photometric data have been transformed from the local instantaneous photometric system to the \citet{1992AJ....104..340L, 2009AJ....137.4186L} standard system via colour equations solved for all frames of each night using a $BVRIgri$ reference sequence located around V3890~Sgr and extracted from APASS DR8 all-sky survey (see \citealt{2014CoSka..43..518H} for details), with colour transformations adopted from \citet{2014AJ....148...81M}.  The same reference sequence has been used at all telescopes on all nights, ensuring a high degree of homogeneity over the entire photometric dataset. The collected photometric data are listed in Table~\ref{tab:UBVRI}, where the quoted uncertainties are the total error budget, which quadratically combines the Poisson uncertainty and the error associated with the transformation to the standard system via the colour equations, which is usually the dominating term.
\begin{table*}
\caption[]{ANS Collaboration $BVRI$ photometry of V3890~Sgr around the
time of the X-shooter observation and at a more recent epoch for comparison. 
The second column provides the HJD$-$2460000.}
\label{tab:UBVRI}
\begin{center}
\begin{tabular}{cccccc}
\hline \hline
\noalign{\smallskip}
Date & HJD & $B$ & $V$ & $R$ & $I$ \\
\noalign{\smallskip}
\hline
\noalign{\smallskip}
2024-08-09.885 & 532.385 & $15.663 \pm0.020$  & $14.813 \pm0.016 $ & $13.780 \pm0.009$  & $12.441 \pm0.008$  \\ 
2024-08-10.871 & 533.371 & $15.667 \pm0.030$  & $14.756 \pm0.024 $ & $13.776 \pm0.026$  & $12.554 \pm0.029$  \\ 
2024-08-20.839 & 543.339 & $15.619 \pm0.086$  & $14.826 \pm0.045 $ & $13.715 \pm0.016$  & $12.321 \pm0.015$  \\ 
2024-08-21.837 & 544.337 & $15.707 \pm0.048$  & $14.799 \pm0.025 $ & $13.732 \pm0.012$  & $12.439 \pm0.015$  \\ 
2024-08-22.827 & 545.327 & $15.524 \pm0.052$  & $14.822 \pm0.025 $ & $13.745 \pm0.013$  & $12.453 \pm0.015$  \\ 
2024-08-24.839 & 547.339 & $15.621 \pm0.021$  & $14.782 \pm0.018 $ & $13.719 \pm0.016$  & $12.354 \pm0.015$  \\ 
2025-06-30.926 & 857.426 & $15.579 \pm0.050$  & $14.710 \pm0.018 $ & $13.745 \pm0.012$  & $12.580 \pm0.016$  \\ 
\hline
\end{tabular}
\end{center}
\end{table*}
\section{Spectral type, extinction and distance}\label{sec:sptype_ext}
\subsection{Spectral type of the red giant}
\begin{figure*}
   \centering

   \includegraphics[width=18cm]{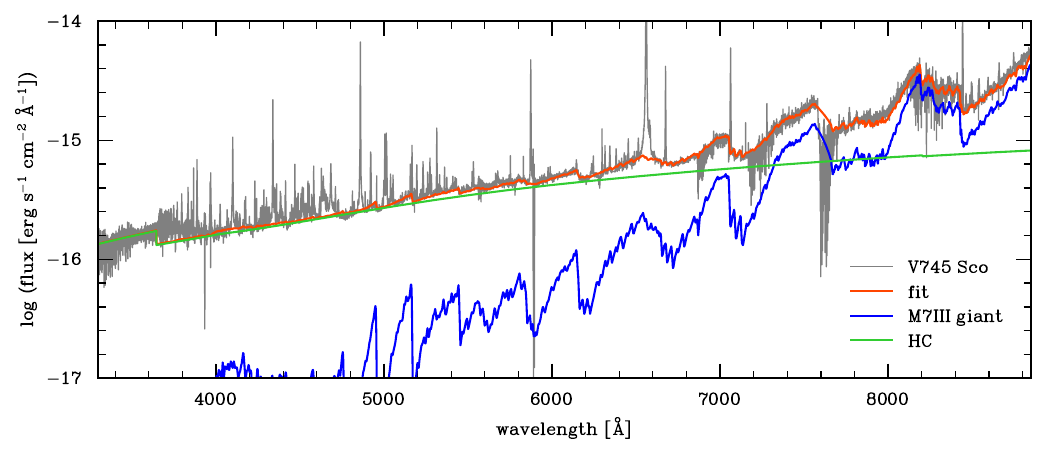}
   \caption{Observational spectrum of V745~Sco in gray, compared to best-fitting spectrum in orange, composed as a sum of the spectra of an M7III giant (blue) and a hot companion (green). The latter is composed from an accretion disk, built from a series of blackbody rings and \ion{H}{i} nebular continuum.}
   \label{fig:spectral_fit_V745Sco}
    \end{figure*}

For the determination of the RG spectral type we roughly follow the procedure in \citet{tha1531}. We make use of the overall shape of the TiO molecular absorption bands that varies strongly among different spectral subclasses by comparing the X-shooter and model spectra. For the RG contribution we adopt standard late-type giant spectra from \citet{Fluks1994}, which are classified according to the Case classification system \citep{1964ApJ...139..190N}. The catalogue accompanying the paper \citep{Fluks1994spectra} includes observed "intrinsic" spectra of late-type giants in the range of 3500--10000\ \AA. Since the Fluks spectra are known to provide an inaccurate representation of the 7200--7500\ \AA \ region \citep[][their Figure 5]{tha1531,Corradi2010Ongoingoutburst}, we substitute them for observational spectra of late-type giant standard stars obtained by the Asiago $1.22\,$m+B\&C telescope\footnote{We use HD 148783 as an M6III and HD 108849 as an M7III standard respectively, the former classified according to \citet{Fluks1994} and the latter according to \citet{CorballyGray2009Stellar}.} below 7500\ \AA. Asiago and Fluks spectra match very closely at wavelengths below 7200\ \AA. The WD and the accretion disk (AD) around it also contribute to the observed system spectra, although in unknown proportions and with unknown spectral shapes. We attempt to model this contribution in several ways, and while the choice of the spectral shape seems crucial when it comes to determining the amount of extinction by the fitting of model spectra (see Sect.~\ref{sec:extincition}), it does not seem important for the determination of the spectral type. We fit the spectra using sigma clipping to avoid influence of strong emission lines, utilizing \texttt{SciPy}'s least squares method. We select the best-fitting spectral class based on visual inspection of the shape of TiO bands redward of 6900 \AA, which vary substantially between spectral classes. We confirm that the selected spectral subclasses correspond to the minima of the root mean square error among the considered templates. Both spectra are adequately reproduced by both M6III and M7III giant templates; however, M7III provides a better match, especially in the case of V745~Sco. We also fit M6.5III spectral subtype, constructed as the geometric mean of M6III and M7III spectra, $F(\lambda)_\mathrm{M6.5III} = \sqrt{F(\lambda)_\mathrm{M6III}\times F(\lambda)_\mathrm{M7III}}$, following the logic from \citet{Fluks1994}, who found that a spectral class can be approximated well as a geometric mean of the neighbouring types (e.g. M7III spectrum can be produced by taking the geometric mean of M6III and M8III spectra). The resulting M6.5III template provides the best match for V3890~Sgr, whereas V745~Sco is still best reproduced by the M7III template (see an example fit for V745~Sco in Fig.~\ref{fig:spectral_fit_V745Sco}). For the remainder of the paper we adopt M7III and M6.5III spectral types for V745~Sco and V3890~Sgr, respectively (we report these, as well as other basic properties of our two systems in Tab. \ref{table:basic_properties}). We estimate the uncertainty in spectral classification to be approximately half a spectral subclass.  
\begin{table}
\caption{Basic properties of V745~Sco and V3890~Sgr.}              
\label{table:basic_properties}      
\centering                                      
\begin{tabular}{c c c c}          
\noalign{\smallskip}
\hline
\hline
\noalign{\smallskip}

Property	   & V745~Sco &   V3890~Sgr & unit  \\

\noalign{\smallskip}
\hline
\noalign{\smallskip}
RG Sp. Type&M7III&M6.5III&\\
\noalign{\smallskip} 
$E(B-V)$   & $0.87\pm0.20$ & $0.59$\,\tablefootmark{(a)}&mag\\
\noalign{\smallskip}
Distance  & $8.5\,^{+3.1}_{-2.3}$ & $7.2\,^{+2.8}_{-2.0}$&kpc\\
\noalign{\smallskip}
$L_\mathrm{UVOT}$ & $56^{+197}_{-43}$&$73^{+67}_{-35}$& $L_{\odot}$\\
\noalign{\smallskip}
$L^{>2000\,\AA}_\mathrm{acc}$ & $340^{+690}_{-240}$ &$360^{+350}_{-180}$ & $L_{\odot}$\\
\noalign{\smallskip}
$\dot{M}_\mathrm{acc}$ &  $\gtrsim2.4\times10^{-8}$ & $\gtrsim2.5\times10^{-8}$ & $M_{\odot}\,\mathrm{yr}^{-1}$\\
\noalign{\smallskip}
$L_\mathrm{IR}$ & $260^{+270}_{-140}$&  & $L_{\odot}$\\
\noalign{\smallskip}
IR type\tablefootmark{1} & $S$+$IR$ & $S$ & \\
\noalign{\smallskip}
RV$_{\mathrm{RG}}$ &  $-124\pm3$ & $ -112\pm3$ & km s$^{-1}$ \\
\noalign{\smallskip}
Orbital period &  &$747\pm4$\,\tablefootmark{(b)} & days\\

\hline                      
\end{tabular}
\tablefoot{\tablefoottext{1}{Per classification in \citet{2019ApJS..240...21A}.} The existence of $S+IR$ type is somewhat uncertain.}
\tablebib{(a)~\citet{2019ATel13069....1M}, (b)~\citet{2021MNRAS.504.2122M}.}

\end{table}

\subsection{Extinction}\label{sec:extincition}
Both systems are located close to the direction of the Galactic centre and at distances of several kiloparsecs, suggesting substantial amounts of interstellar extinction.
\subsubsection{V745~Sco} \label{sec:V745_ext}
Previous studies report varying findings for the extinction towards V745~Sco. \citet{2024MNRAS.534.1227M} report $E(B-V) = 0.69\pm0.20\ \mathrm{mag}$ based on the equivalent width (EW) of absorption features of several diffuse interstellar bands (DIBs). \citet{2018MNRAS.476.4162O} find $E(B-V) = 1.0\pm0.2\ \mathrm{mag}$ by calculating the dimming of the outburst maximum brightness apparent magnitude caused by interstellar extinction, while \citet{2018ApJS..237....4H} discuss extinction to a greater extent, deriving $E(B-V)=0.9\pm0.2\ \mathrm{mag}$ from hydrogen column density measured by X-ray post-outburst spectral fitting, and in the end adopting $E(B-V) = 0.70$ based on the Galactic dust extinction by \citet{2011ApJ...737..103S}.
\par
We first attempt to derive the extinction from fitting the model spectrum. However, we find that the fitting is unreliable, as the choice of the WD+AD contribution model greatly influences the derived $E(B-V)$ value. Adopting a bremsstrahlung model with temperature as a free parameter, we derive $E(B-V)\sim1.1\ \mathrm{mag}$, placing it at the upper end of the values reported in the literature. Alternatively, we model the accretion disk spectrum as a sum of blackbody emitting rings, with the temperature of the rings being dependent on their distance from the WD \citep[see][Ch.~18, Eq.~19]{2017imas.book.....C} and add the spectrum of continuously-emitting \ion{H}{i} at $10000\,$K \citep[see][Ch.~4, Table~4.7]{2006agna.book.....O}. Such an approach yields $E(B-V)\sim0.6\ \mathrm{mag}$, which is at the opposite, lower, end of the values reported in the literature. 
\par
Using the \ion{Na}{i}~D doublet as a proxy for extinction proves to be unfeasible, as we are unable to separate the interstellar and stellar absorption features at this resolving power. We also attempt to use the EW of the \ion{K}{i} line at $7699\ $\AA\ to estimate the $E(B-V)$ value using the relation in \citet{1997A&A...318..269M}. In this process we struggle to determine the stellar continuum value, so we are only able to constrain the extinction to $0.65\ \mathrm{mag}<E(B-V)<1.1\ \mathrm{mag}$.
\par
We also calculate the reddening by comparing the $J-K_{\rm s}$ colour of V745~Sco to that of a typical M7III bulge giant. A distinction between bulge and field giants is necessary because \citet{1987ApJ...320..199F} reported that typical bulge M-giants appear to be bluer and fainter, compared to those in the Solar neighbourhood. We believe that comparison of our systems to typical bulge M-giants makes sense because i) based on the coordinates and the large distance estimates to our two objects (see Sect. \ref{sec:distances} for distances derived in previous studies), it seems reasonable for the two systems to belong to the bulge, and ii) as stated by \citet{1992A&A...255..171W}, the majority of giants in symbiotic stars exhibit infrared colours similar to those of bulge stars. Using the bulge M-giant list from \citet{1987ApJ...320..199F} (see Appendix~\ref{sec:app_Bulge} for details), we calculate $J-K_{\rm s}=1.28\pm{0.11}$ as the mean\footnote{We use mean instead of median colours because neither appears strongly preferred from inspection of the distributions, and the uncertainties of the mean values can be propagated more naturally.} colour of a typical M7III bulge giant. Comparing this value to the $J-K_{\rm s}$ colour of V745~Sco and following the extinction conversion relations from \citet{2003A&A...401..781F}, we calculate $E(B-V) = 0.87\pm0.20$. The quoted uncertainty is obtained by propagating the uncertainties on the mean $J-K_{\rm s}$ colour for a bulge M7III giant, and the uncertainties of measured V745~Sco $J$ and $K_{\rm s}$ magnitudes. Additionally, we note that there are two contributions to the systematic uncertainty, stemming from possible misclassification of the giant as an M7III type (rather than an M6.5III type) and/or from the assumption that the giant resembles bulge late-type giants rather than the ones in the Solar neighbourhood. We discuss them in Appendix~\ref{app:systematic_uncertainties} and adopt $E(B-V) = 0.87\pm0.20$ for the remainder of this paper. 

\subsubsection{V3890~Sgr}
A detailed review of the previous investigations of reddening to V3890~Sgr is given in \citet{2020MNRAS.499.4814P}. They report published estimates of $E(B-V)$ spanning the range from just below 0.5 to approximately 1.1 mag. Since then, most of the studies have adopted $E(B-V) = 0.59\ \mathrm{mag}$ referencing \citet{2019ATel13069....1M}, who report the value as a mean of values derived from the EWs of \ion{Na}{i}~D doublet and DIB absorption features. \citet{2022MNRAS.517.6064K} obtain the value $E(B-V)=0.4\ \mathrm{mag}$ from spectral fitting, although they do not take the contribution from the WD or the AD into account, commenting that the $E(B-V)$ value could thus be underestimated.\par
Our estimation of the extinction from spectral fitting appears similarly unreliable. If we model the WD+AD contribution as bremsstrahlung, we derive $E(B-V)\sim0.5\ \mathrm{mag}$, while for a ring blackbody + continuum emission model we derive $E(B-V)\sim0.3\ \mathrm{mag}$. In both cases, the values are lower than those reported in other studies, although the discrepancy is not as large as in the case of V745~Sco.\par
The interstellar and stellar components of the \ion{Na}{i}~D doublet are again unresolved, while the \ion{K}{i} 7699\ \AA\ line EW constrains the reddening to $0.45\ \mathrm{mag}< E(B-V)<0.65\ \mathrm{mag}$ following \citet{1997A&A...318..269M}.\par 
The assumption of bulge membership for this system is further supported by \citet{2021MNRAS.504.2122M}, who determined that V3890~Sgr belongs to the bulge, based on the systemic radial velocity of V3890~Sgr not being consistent with Galactic rotation at any point along its line of sight. We thus also compare the V3890~Sgr colour to the $J-K_{\rm s}=1.20\pm{0.14}\ \mathrm{mag}$ mean\footnotemark[\value{footnote}] intrinsic bulge M6.5III giant colour and recover $E(B-V) = 0.64\pm0.25\ \mathrm{mag}$, again noting the presence of an unknown systematic uncertainty. This value agrees well with $E(B-V)=0.59$ \citep{2019ATel13069....1M}, which we consider the most robustly determined estimate and we hence adopt for the remainder of this paper. 
\subsection{Distance}\label{sec:distances}
\subsubsection{V745~Sco}
The distance to V745~Sco that has often been adopted in the past is $7.8\pm1.8\,\mathrm{kpc}$ from \citet{2010ApJS..187..275S}. \citet{2024MNRAS.534.1227M} argue that the assumption of orbital period that is used to determine this distance is incorrect and derive the distance to be $8.2^{+1.2}_{-1.0}\ \mathrm{kpc}$, based on a simulation of apparent magnitudes of novae in a realistic model of the Galaxy. Both of these values are consistent with the 8 kpc estimate by \citet{1990MNRAS.246...78S} and the $8.0\pm0.6\ \mathrm{kpc}$ distance reported by \citet{2022MNRAS.517.6150S}.\par 
The uncertainty on the Gaia Early Data Release 3 \citep[Gaia EDR3;][]{2016A&A...595A...1G,2021A&A...649A...1G} parallax ($0.0999\pm0.0764\ \mathrm{mas}$) is too large to be safely used in estimating the distance to the source, indicating only that the source lies at a distance of several kiloparsecs. 3-D maps of Galactic extinction \citep{2006A&A...453..635M} only provide us with the lower limit of the distance, fixing it to $>4\ \mathrm{kpc}$, as the reported extinction is asymptotic beyond this distance. Following the arguments presented in Sect.~\ref{sec:V745_ext}, we again use the bulge M-giant sample of \citet{1987ApJ...320..199F}. We derive the characteristic dereddened $K_{\rm s}$-band apparent magnitude for an M7III type giant in the galactic bulge $K_{\rm s}=7.76\pm0.67$. Comparing this value with the V745~Sco 2MASS K$_{\rm s}$ magnitude, assuming a bulge distance of $8\ \mathrm{kpc}$, and accounting for the reddening, we can estimate the distance to V745~Sco to be $8.5^{+3.1}_{-2.3}\,\mathrm{kpc}$, where the uncertainty in distance estimate reflects the uncertainties in derived $E(B-V)$ of V745~Sco, the uncertainty of V745 Sco 2MASS $K_\mathrm{s}$ magnitude, and the mean $K_{\rm s}$ of M7III bulge giants. A potentially significant systematic uncertainty is again present, arising from the adopted spectral classification and from the assumption that the giant resembles typical bulge giants. We adopt the distance to be $8.5^{+3.1}_{-2.3}\,\mathrm{kpc}$ for the remainder of this paper.
\subsubsection{V3890~Sgr}
The distance to V3890~Sgr has historically been less well constrained. \citet{1992A&A...265...71G} estimated lower and upper distance limits of 2.6 kpc and 5 kpc, respectively, based on assumptions of the maximum absolute magnitudes of recurrent novae and the $E(B-V)$ value. \citet{2010ApJS..187..275S} derives the distance to be $7.0 \pm 1.6\ \mathrm{kpc}$ using brightness, temperature and size estimates of the RG. Their earlier paper \citep{2009ApJ...697..721S} gives the distance estimate of $6 \pm 1\ \mathrm{kpc}$, as a mean of three individual methods, while their most recent paper \citep{2022MNRAS.517.6150S} derives the distance to be $8.5\pm0.5\ \mathrm{kpc}$ by combining posteriors of multiple methods. \citet{2021MNRAS.504.2122M} adopt the distance to V3890~Sgr to be $9\ \mathrm{kpc}$ after considering multiple procedures that yield similar values. Most recently, \citet{2026MNRAS.552g1851M} used VLBI imaging of the 2019 outburst to calculate the expansion parallax distance of $6.8\ \mathrm{kpc}$. They considered distances ranging from $5.9\ \mathrm{kpc}$ to $11.4\ \mathrm{kpc}$, depending on assumptions on the expansion velocity of the outburst. \citet{2019ATel13069....1M} estimate the lower limit of the distance to be $\sim4.5\,\mathrm{kpc}$ from the measurements of the interstellar extinction to the star in combination with (3D) maps of interstellar dust \citep{2014A&A...561A..91L,2019ApJ...887...93G}.
\par 
The Gaia EDR3 parallax is again too uncertain ($0.0484\pm0.0453\,\mathrm{mas}$) to derive anything but a several kiloparsec distance to the system. The Galactic extinction maps of \citet{2006A&A...453..635M} only give the lower limit of the distance at $4\,\mathrm{kpc}$. Using the same arguments as for V745~Sco, we derive a characteristic dereddened apparent magnitude of $K_{\rm s}=8.22\pm0.71$ for an M6.5III bulge giant, which yields a distance estimate of $7.2^{+2.8}_{-2.0}\,\mathrm{kpc}$ to V3890~Sgr. In this estimate we use $E(B-V)=0.59$ from \citet{2019ATel13069....1M}, so the uncertainty on distance only propagates the uncertainty on the typical bulge giant $K_{\rm s}$ magnitude and the measured V3890~Sgr $K_{\mathrm{s}}$ magnitude. We again note the discussed systematic uncertainties. We adopt a distance of $7.2^{+2.8}_{-2.0}\,\mathrm{kpc}$ for the remainder of this paper. 

As a consistency check, we note that the derived distances are consistent with the giants belonging to the bulge, but given that they are derived based on this assumption in the first place, they should not be interpreted as independent measurements of the systems' position.  
\section{Accretion luminosity and IR excess}\label{sec:SED}

We compare the dereddened spectral energy distributions (SEDs) of our objects to those of typical bulge late-type giants of the appropriate spectral type (M6.5III/M7III) and use the differences to analyse the accretion signatures.
\subsection{Model and object SED construction}
The SED of V745~Sco is constructed using photometric measurements from several sources. For $B$, $V$, $R$ and $I$ bands we use the AAVSO observations that we used for fluxing the X-shooter spectrum. The $J$, $H$ and K$_{\rm s}$ magnitudes are obtained from the 2MASS Point Source Catalogue \citep{2003yCat.2246....0C}. The $W1$ and $W2$ are taken as NEOWISE \citep{2011ApJ...731...53M} averages since 2022, while $W3$ and $W4$ are taken from the AllWISE \citep{2010AJ....140.1868W,2014yCat.2328....0C} part of the survey. The V3890~Sgr SED is constructed similarly, with the ANS collaboration observations providing the $B$, $V$, $R$ and $I$ magnitudes, and the NEOWISE $W1$ and $W2$ bands being averaged since the outburst in 2019. While most of the observations were conducted quasi-simultaneously, 2MASS, and AllWISE data were collected 25 and 15 years before the epoch of the X-shooter pointings, respectively. Because both of our sources were in deep quiescence at those times (preceding outbursts in 1989 and 1990 for V745~Sco and V3890~Sgr, respectively) we consider the resulting systematic uncertainty to be small compared to the effects discussed below. For V745~Sco this is further supported by the fact that AllWISE $W1$ and $W2$ magnitudes closely match NEOWISE averages since 2022, so it is reasonable to assume that $W3$ and $W4$ would be similar as well. We deredden the photometric data according to prescriptions from \citet{2003A&A...401..781F} and \citet{2023ApJS..264...14Z}, using $E(B-V)$ values from Sect.\ref{sec:extincition}.
\par 
To construct reference (intrinsic) giant SED for comparison, we again rely on the list of bulge giants from \citet{1987ApJ...320..199F}. We construct the typical colours by again taking the mean colours of giants within a spectral type. We use OGLE $V$ and $I$ band data, 2MASS $J, H $ and $K_{\rm s}$ magnitudes, and AllWISE $W1$, $W2$, $W3$ and $W4$ magnitudes (see Appendix~\ref{sec:app_Bulge}). To acquire $B$ band colour we relied on \citet{1970A&A.....4..234F}, who determine $B-V=1.5$ for M6III and M7III giants. We estimate $U-B\approx1.1$ as a value that appears reasonable based on a collection of different M6III/M7III photometric measurements \citep[e.g.][]{1970ApJ...162..217L,1970A&A.....4..234F,Fluks1994}. The final connection is made by using IUE Spectral Atlas \citep{1982ESASP.182...25W} to estimate the flux density ratios of $UVW1$ to $U$ (1:8) and $UVW2$ to $U$ (1:12). These estimates are primarily intended to provide a reference UV SED for qualitative comparison. The inferred UV excess is sufficiently large that the precise choice of UV colours does not affect our conclusions.
\par
\begin{figure}
  \resizebox{\hsize}{!}{\includegraphics{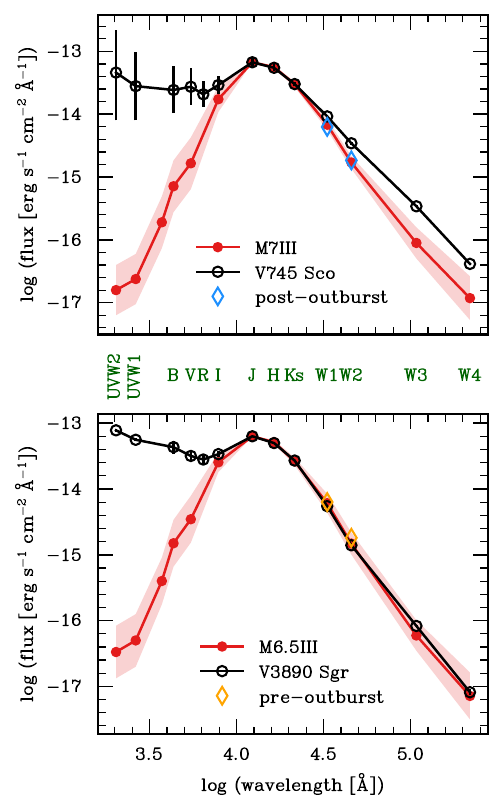}}
  \caption{The comparison of the dereddened SEDs of V745~Sco and V3890~Sgr to those of typical late-type bulge giants. Typical giant SEDs are scaled such that they match object SEDs in $J$ band. Top: comparison of V745~Sco (black) to a typical M7III giant (red). Blue diamonds indicate state of the system immediately after the outburst (see Sect.~\ref{sec:IR_excess}). Bottom: comparison of V3890~Sgr (black) to a typical M6.5III giant (red). Orange diamonds indicate the state of the giant immediately before 2019 outburst. Passband names between the two panels are horizontally aligned to the respective datapoints. Errorbars in the upper plot include uncertainty on $E(B-V)$ and individual observations, while only uncertainty on individual observations is present in the bottom. Light red band indicates uncertainty of our model colours.}
  \label{fig:seds}
\end{figure}

We compare the SEDs of our two sources in Fig.~\ref{fig:seds}. It is clear that both of the sources exhibit a strong UV excess. V745~Sco exhibits IR excess as well, which is, however, not present in the SED of V3890~Sgr. We discuss and quantify these features in the next paragraphs.

\subsection{UV excess}\label{sec:UV_excess}
The presence of ongoing accretion in both systems is evident from the pronounced UV excess. Integrating the excess flux density provides a lower-limit estimate of the accretion luminosity, a value that is often used to quantify the level of accretion in SySts. We calculate two values. The first, $L^{>2000\,\AA}_\mathrm{acc}$, is obtained by integrating the area between the black and red curves in Fig.~\ref{fig:seds} at wavelengths shortward of the midpoint of the 2MASS $J$ band ($\lambda < 12318\ \text{\AA}$), where the accretion-related excess becomes noticeable. We obtain the other by integrating across the Swift/UVOT $UVW1$ and $UVW2$ bands ($L_\mathrm{UVOT}$), as that is a value which is often reported in the literature and is as such useful for comparison with other SySts/RNe\footnote{The value reported in other literature typically includes $UVM2$ band flux in addition to $UVW1$ and $UVW2$ band fluxes. As we lack $UVM2$ photometry, we instead interpolate and integrate from $UVW1$ to $UVW2$. Our calculation therefore ignores the potential drop in the $UVM2$ flux observed in SU~Lyn \citep{2016MNRAS.461L...1M} and THA 15-31 \citep{tha1531}. This drop is most likely caused by the $2175\ \text{\AA}$ extinction bump.}.
\newline Assuming solar luminosity $L_\odot = 3.828~\times10^{33}\ \mathrm{erg}\ \mathrm{s}^{-1}$, we calculate the accretion luminosities of V745~Sco to be
\begin{equation*}
    L^{\mathrm{V745~Sco}}_\mathrm{acc,\ >2000\,\AA} = 340^{+690}_{-240}\ L_\odot
\end{equation*}
and 
\begin{equation*}
    L^{\mathrm{V745~Sco}}_\mathrm{UVOT} = 56^{+197}_{-43}\ L_\odot, 
\end{equation*}
whereas for V3890~Sgr we calculate 
\begin{equation*}
    L^{\mathrm{V3890~Sgr}}_\mathrm{acc,\ >2000\,\AA} = 360^{+350}_{-180}\ L_\odot
\end{equation*}
and 
\begin{equation*}
    L^{\mathrm{V3890~Sgr}}_\mathrm{UVOT} =73^{+67}_{-35}\ L_\odot. 
\end{equation*}
The reported uncertainties were determined by propagating the uncertainties of individual photometric measurements, model SED, as well as our distance and extinction estimates, and are therefore smaller for V3890~Sgr, for which no uncertainty estimate is available for the adopted $E(B-V)$. Within these admittedly large uncertainties, the accretion luminosities of the two objects are consistent with each other. Integrating the areas under the red curves in Fig.~\ref{fig:seds} returns RG luminosities of $L^\mathrm{V745\ Sco}_\mathrm{RG} = 2300\ L_\odot$ and $L^\mathrm{V3890\ Sgr}_\mathrm{RG} = 1600\ L_\odot$, corresponding to bolometric magnitudes of $M^\mathrm{V745\ Sco}_\mathrm{bol} = -3.65$ and $M^\mathrm{V3890\ Sgr}_\mathrm{bol} = -3.25$. As a sanity check we note that such values match well with the expected bolometric magnitudes of bulge late-type giants \citep{1982ApJ...259L...7F}. 

\subsection{IR excess}\label{sec:IR_excess}

Characteristics of the infrared excess have traditionally been used to classify SySts into $S$-type (stellar) and $D$-type (dusty) systems \citep{1974MNRAS.167..337A}. The SEDs of $S$-type systems are well reproduced by a single $\sim3000\ $K blackbody corresponding to the red giant, whereas $D$-type systems require an additional infrared component, typically consistent with dust radiating at $\sim1000\ $K. \citet{1982ASSL...95...27A} introduced the $D'$ class, whose SEDs match an F- or G-type giant and IR excess at $2.2-3.6\ \mu\mathrm{m}$ (around $K_\mathrm{s}$ and W1 bands). More recently, \citet{2019ApJS..240...21A} identified a new class, denoted $S+IR$, in which excess emission is only present in the WISE W3 and W4 bands. These objects can be described by the combination of an M-type giant SED and a cooler infrared component with a characteristic temperature of approximately $400\ $K.

\begin{figure}    
   \resizebox{\hsize}{!}{\includegraphics{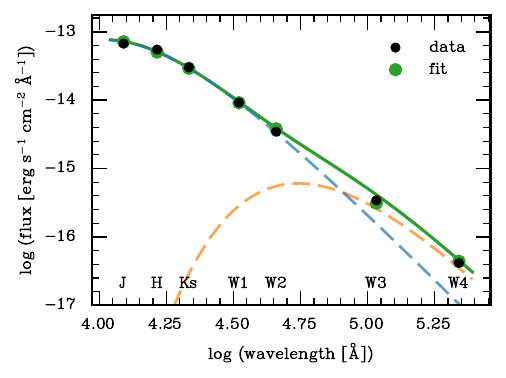}}
   \caption{A fit of IR colours of V745~Sco with two blackbody spectra at different temperatures. Black points show observational data. Blue and orange dashed lines show individual blackbody spectra for $T_\mathrm{BB}\sim2700\ $K and $T_\mathrm{BB}\sim530\ $K, respectively. Full green line shows the sum of the two, while green dots show passband-integrated green line.}
   \label{fig:two_planck_V745}

\end{figure}

Analysing the two sources in such a manner, we classify V3890~Sgr as an $S$-type system, being well fitted with a single Planck SED with $T_\mathrm{BB}\sim3300\,\mathrm{K}$, and V745~Sco as an $S$+$IR$-type system, the SED of the latter being well described with two Planck SEDs, one with $T_\mathrm{BB}\sim 2700\,\mathrm{K}$ and the other with $T_\mathrm{BB}\sim530\,\mathrm{K}$ (Fig.~\ref{fig:two_planck_V745}). 

\begin{figure}
  \resizebox{\hsize}{!}{\includegraphics{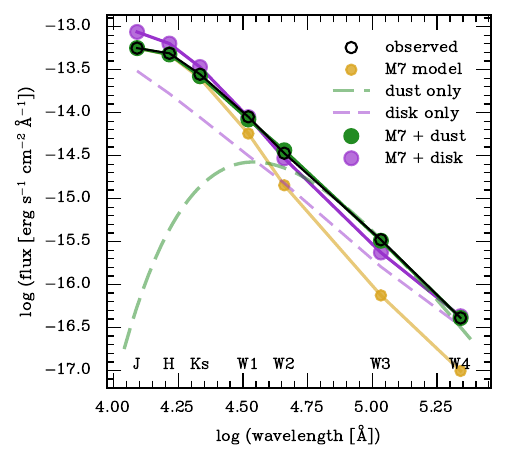}}
  \caption{Comparison of V745~Sco IR-excess modelling via dust or disk. Black points show observational data, gold points show our model M7III giant with no companion. Dust model contribution is plotted as green dashed line, with green points showing the band-integrated combined dust and giant model. Disk model contribution is shown in purple, with the purple points showing band-integrated combined disk and giant model.}
  \label{fig:IR_excess_modeling}
\end{figure}

We point out that the resulting temperatures used to describe the near-IR part of the SED, dominated by the RG, should not be mistaken for effective temperatures of the giants. In fact, \citet{2019ApJS..240...21A} derive a relation between $T_\mathrm{eff}$ of the giant and the $T_\mathrm{BB}$ that best fits its SED. Using the relation, we derive $T_\mathrm{eff}^\mathrm{V745~Sco} \sim3500\,\mathrm{K}$ and $T_\mathrm{eff}^\mathrm{V3890~Sgr} \sim3900\,\mathrm{K}$, both significantly higher than those expected for a standard M-giant of appropriate spectral subtype. While this could indicate that the cool giants in our sources are indeed hotter (and thus bluer) than their counterparts in the Solar neighbourhood, we are using $E(B-V)$ values which are derived based on this assumption in the first place, nearing a circular argument. Similarly, the $T_\mathrm{BB}$, which is used to fit the excess in W3 and W4 bands, should not be interpreted as the temperature of the radiating dust. In fact, comparing the upper plot in Fig.~\ref{fig:seds} to Fig.~\ref{fig:two_planck_V745}, one can see that if we model the SED as a combination of two Planck spectra, the IR excess only begins to show in W3 and W4 bands, whereas if we compare V745~Sco to a model RG, the excess is evident in all four WISE passbands. Fitting the residuals from Fig.~\ref{fig:seds} with a Planck SED (shown in Fig.~\ref{fig:IR_excess_modeling}) we get a temperature of $\sim780\,\mathrm{K}$ instead, which we believe to be more indicative of the actual temperature of the emitting dust. While rather high, it remains below the condensation temperature of silicate dust grains \citep{1999MNRAS.304..389S}. Integrating between the black and gold points in Fig.~\ref{fig:IR_excess_modeling}, we find the excess IR luminosity in V745~Sco to be $L_\mathrm{IR}^\mathrm{V745~Sco}=260^{+270}_{-140}~L_\odot$, where we again propagate the uncertainties on individual photometric measurements, model SED, extinction and distance.
\subsubsection{IR contamination}
Because both objects are located in relatively crowded fields, we assessed the possibility that the AllWISE photometry is contaminated by nearby sources, particularly in the W4 band, whose angular resolution is approximately 12\arcsec\ \citep{2010AJ....140.1868W}. We identify two additional 2MASS sources within a radius of 12\arcsec\ around V3890~Sgr and five around V745~Sco. 

The two sources near V3890~Sgr are 4.9 and 5.0\,mag fainter in $K_{\rm s}$ and are respectively 0.5 and 0.3 mag bluer in $J-K_{\rm s}$. Their combined flux in the $K_{\rm s}$ band amounts to only $\sim2\%$ of that of V3890~Sgr. Since both objects are also bluer than V3890~Sgr, their contribution is expected to be even smaller at longer wavelengths. We therefore consider their effect on the WISE photometry negligible. 

The situation is somewhat less straightforward for V745~Sco. The five neighbouring sources are between 2.8 and 4.0~mag fainter in $K_{\rm s}$, and their combined $K_{\rm s}$-band flux corresponds to approximately 24\% of that of V745~Sco. However, all five sources are bluer by 0.4--1.3 mag in $J-K_{\rm s}$, suggesting that their relative contribution decreases towards longer wavelengths. As an illustrative estimate, if the differences in $K_{\rm s}-W4$ were comparable to those in $J-K_{\rm s}$, the combined contaminating flux would correspond to $\sim16\%$ of the flux of V745~Sco in W4. If the neighbouring stars were even bluer in $K_{\rm s}-W4$, the contamination would be correspondingly smaller. 

Visual inspection of the AllWISE images reveals a slight asymmetry of the source profile in W1 that may be caused by nearby stars. A weaker effect may also be present in W2, but no obvious distortion is visible in W3, despite its angular resolution being similar to that of W1 and W2 ($\sim6\arcsec$). Furthermore, the catalogued WISE magnitudes are derived from PSF fitting rather than aperture photometry, making them less sensitive to contamination from nearby sources. Taken together, these considerations suggest that source blending is unlikely to be the dominant factor affecting the WISE photometry of V745~Sco. 

The strongest evidence against significant source blending for the case of V745~Sco is the temporal evolution of the WISE photometry. The W1 and W2 excess disappears following the 2014 outburst and subsequently reappears over several years (see Fig.~\ref{fig:seds} and Fig.~\ref{fig:neowise_evolution}). Since contaminating field stars would contribute a constant flux at all epochs, they cannot account for the observed behaviour. While contamination at approximately 10--15\% level cannot be ruled out, it is not capable of explaining the observed mid-infrared excess in full, and does not affect our conclusions qualitatively.
\subsubsection{Source of IR excess in V745~Sco}
\begin{figure}
  \resizebox{\hsize}{!}{\includegraphics{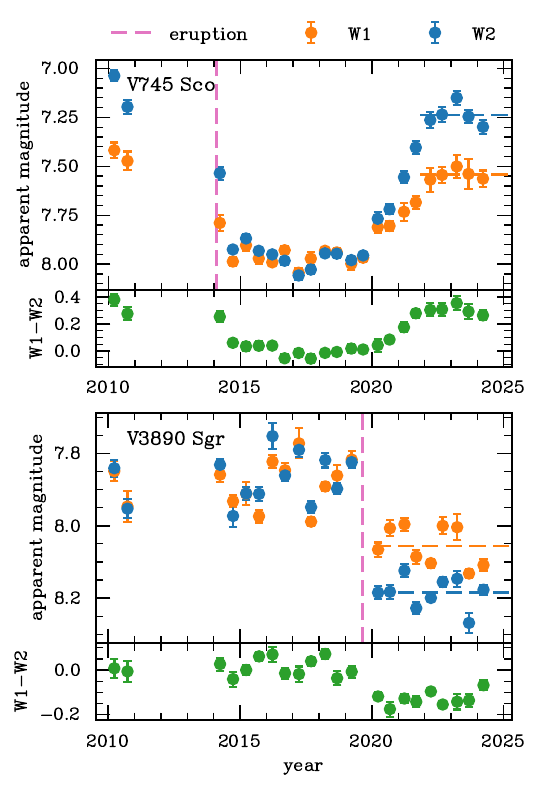}}
  \caption{WISE W1 and W2 epoch photometry of V745~Sco (top) and V3890~Sgr (bottom). Earliest two datapoints were collected during the WISE cryogenic phase (AllWISE), while the rest were collected during the NEOWISE part of the mission. Pink vertical lines mark the first detection of the outburst of the corresponding star. Green dots show the evolution of the $W1-W2$ colour. Dashed horizontal lines show averages used in SED construction.}
  \label{fig:neowise_evolution}
\end{figure}

As hinted by the traditional classification of SySts into stellar and dusty types, IR excess is typically attributed to thermal emission of cold circumstellar dust, originating mostly from condensed stellar wind of the RG. \citet{2025A&A...701A.176M} have, however, confirmed that the accretion disk in T~CrB is extended up to $\sim58\ R_\sun$. Such an extended disk may have outer regions cool enough to contribute significantly to the observed IR excess. Both of these interpretations are consistent with the epoch photometric data of V745~Sco from All+NEOWISE surveys, shown in the top panel of Fig.~\ref{fig:neowise_evolution}, which shows a decrease in brightness after the outburst (with the first post-outburst datapoint probably still being influenced by the outburst itself), followed by 5--6 years of quiescence, during which it matches a model M7III bulge giant (see Fig.~\ref{fig:seds}), before increasing to pre-outburst brightness over a period of approximately 2 years. Such a photometric evolution could be explained well either by a dusty envelope, which is blown away during a nova outburst, before reforming from the condensed wind of the RG in the following years, or by an accretion disk, which would be flushed onto the WD, causing the outburst, or be destroyed in the outburst itself, before slowly reforming as the mass transfer resumes after the outburst. 
\par
Assuming that a dusty envelope is the cause of the IR excess, we can estimate the mass of the emitting dust. We consider two similar, yet separate approaches. Modelling the dust as an optically thin cloud of silicate dust grains radiating as blackbodies at a given temperature $T_d$, we use the standard equation \citep[e.g.][]{1988AJ.....95.1817K,2026PASP..138d3001S} to calculate the mass of the radiating dust
\begin{equation}
    M_\mathrm{dust} = \frac{F_{\lambda}d^{2}}{B_{\lambda}(\lambda,T_\mathrm{d})\kappa_{\lambda}}, 
\end{equation}
where $F_{\lambda}$ is the flux density of the source at wavelength $\lambda$, $d$ the distance to the source, $B_{\lambda}(\lambda,T_\mathrm{d})$ Planck function for a dust of temperature $T_\mathrm{d}$, evaluated at $\lambda$ and, $\kappa_{\lambda}$ is the dust mass opacity for a given size of grains. We evaluate at $\lambda=10\, \mu\mathrm{m}$. This seems like a good choice because it is firmly within our wavelength range, and, more importantly, because the dust mass opacity for silicate grains is $\kappa_{10\ \mu\mathrm{m}}\sim1000\ \mathrm{cm^{2}/g}$, independently of their size \citep{2018A&A...617A.124Y}. Using $T_\mathrm{d}=780\ \mathrm{K}$ (see Sec.~\ref{sec:IR_excess}), we find the mass of the dust that is emitting the IR excess in V745~Sco to be $M_\mathrm{dust}=1.4\times10^{-9}\ M_\odot$. 

If we instead assume that the emitting dust is similar to that observed in condensed nova ejecta, i.e. it is composed of spherical carbon dust grains with radii $a<1\ \mathrm{\mu m}$, temperature $T_\mathrm{d}<1000\ \mathrm{K}$, that the density is $\rho\sim2.3\ \mathrm{g\ cm^{-3}}$ and that the Planck emission cross section for dust with such properties is $Q_\mathrm{e}=0.01aT_{d}^{2}$ \citep[e.g.][]{1974ApJS...28..397G,1996ApJ...470..577M,2008A&A...492..145M}, we can calculate the mass of the emitting dust from the equation: 
\begin{equation}
    M_\mathrm{dust}=1.17\times10^{6}\rho T_\mathrm{dust}^{-6}L_\mathrm{IR}.
\end{equation}
In this equation $L_\mathrm{IR}$ is the luminosity of the infrared excess in units of $L_\odot$. We use $T_\mathrm{d}=780\ \mathrm{K}$ and $L_\mathrm{IR}^\mathrm{V745~Sco}=260~L_\odot$ (see Sec.~\ref{sec:IR_excess}) to estimate the mass of the emitting dust to be $M_\mathrm{dust}=3\times10^{-9}\ M_\odot$. Both methods yield dust mass estimates of order $10^{-9}\ M_\odot$. 

For the accretion disk to be the source of the IR excess, the outer layers have to be cool enough to peak in the WISE passbands. Modelling the accretion disk SED as the sum of concentric blackbody rings (see Sect.~\ref{sec:V745_ext}), and assuming $M_{WD}=1.35\ \mathrm{M}_\odot$, $R_{WD}=2100\ \mathrm{km}$ and mass transfer rate $\dot{M}_\mathrm{acc}=2.5\times10^{-8}\ M_\odot\ \mathrm{yr}^{-1}$ (see Sect. \ref{sec:mass_transfer_rate}), we find that the accretion disk would need to extend to about $\mathrm{R}=40\ \mathrm{R}_\odot$ for the temperature to drop to $T=400\ \mathrm{K}$, which is below outer radii of accretion disks in other symbiotic recurrent novae \citep[e.g.][]{2025A&A...701A.176M}. Although a sufficiently extended accretion disk can in principle reach temperatures capable of producing WISE-band excess emission, we were unable to reproduce the observed SED without simultaneously introducing excess to predicted 2MASS fluxes. Such attempts also find the outer radius of the accretion disk to extend to unexpectedly large $r_\mathrm{out}\sim130\ R_\odot$. This may indicate that a dusty circumstellar component provides a slightly more natural explanation for the observed infrared excess. However such a mismatch alone could easily be attributed to the simplicity of our model. We are assuming the disk to be perfectly flat and thin, and are not accounting for irradiation of the disk by the WD, by the hotspot, or by its own hot inner parts. Both dust and disk thus remain viable as the source of observed IR excess. 

\begin{figure}
  \resizebox{\hsize}{!}{\includegraphics{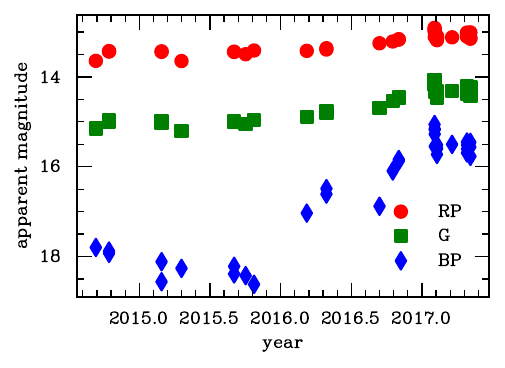}}
  \caption{Gaia DR3 epoch photometry of V745 Sco. Red circles, green squares and blue diamonds correspond to RP, G and BP bands, respectively.}
  \label{fig:gaia}
\end{figure}

It is also interesting to compare the post-nova evolution of the system in WISE bands to that observed in Gaia DR3 epoch photometry \citep[][]{2021A&A...649A...3R,2023A&A...674A...1G} in Fig.~\ref{fig:gaia}, which shows the reformation of the inner layers of the accretion disk beginning between 2015.8 and 2016.2 and completing by 2017.1. Should the increase in the WISE $W1$ and $W2$ bands correspond to the reformation of the outer parts of the disk, the 5 year delay between the reformation of inner and outer parts of the accretion disk would indicate some type of inside-out disk reformation. If we assume that the reformation timeline is directly correlated to the orbital period of the system, being primarily driven by viscous processes, we can compare it to the timeline of disk reformation observed in RS~Oph, in which disk is reformed in approximately half of its orbital period. If such a scaling was applicable, the observed timescale would be consistent with an orbital period of several years. Given the large number of assumptions involved, this estimate should be regarded as highly speculative.

\section{Mass transfer rate and inter-outburst period}\label{sec:mass_transfer_rate}
Assuming that the entire gravitational potential energy of the infalling mass is eventually released as accretion luminosity, we calculate the mass transfer rate $\dot{M}_\mathrm{acc}$ from the bolometric accretion luminosity of the systems $L^\mathrm{bol}_\mathrm{acc}$, using the equation
\begin{equation}\label{eq:masstransferrate}
    \dot{M}_\mathrm{acc} = \frac{L^\mathrm{bol}_\mathrm{acc}R_\mathrm{WD}}{GM_\mathrm{WD}}, 
\end{equation}
where $R_\mathrm{WD}$ and $M_\mathrm{WD}$ are the radius and the mass of the WD, and $G$ is the gravitational constant. We adopt the mass of the WD in both objects to be $M_\mathrm{WD} = 1.35\ M_\odot$, a value that is commonly adopted for all four symbiotic RNe, and is also supported by observational evidence \citep[e.g.][for V745~Sco and V3890~Sgr, respectively]{2015MNRAS.454.3108P,2021MNRAS.504.2122M}. Mass-radius relations for high-mass WDs from \citet{2022A&A...668A..58A} and \citet{2023MNRAS.523.4492A} give $R_\mathrm{WD}$ between $1500$ and $2200$ km for a $M_\mathrm{WD} = 1.35\ M_\odot$, depending on whether an ONe or a CO WD is assumed, and whether general relativity is accounted for. As there are claims in the literature that both of our objects harbour a CO WD \citep[][for V745~Sco and V3890~Sgr, respectively]{2017MNRAS.464.5003O,2021MNRAS.504.2122M}, we adopt $R_\mathrm{WD} = 2100\ \mathrm{km}$ as a working assumption. 

A simple way of estimating the lower limit of the mass transfer rate is to substitute true bolometric accretion luminosity, which we cannot calculate due to lack of wavelength coverage shortwards of UVW2, with $L^{>2000\,\AA}_\mathrm{acc}$, which we calculated in Sect.~\ref{sec:UV_excess}. Plugging our WD parameters and $L^{>2000\,\AA}_\mathrm{acc}$ into Eq.~\ref{eq:masstransferrate}, we calculate this lower limit in our objects to be $\dot{M}_\mathrm{acc}\gtrsim2.4\times10^{-8}\ M_{\odot}\ \mathrm{yr}^{-1}$ and $\dot{M}_\mathrm{acc}\gtrsim2.5\times10^{-8}\ M_{\odot}\ \mathrm{yr}^{-1}$, for V745~Sco and V3890~Sgr, respectively, around the epoch of the X-shooter observations. Since the spectral energy distribution likely peaks at wavelengths shorter than $UVW2$, the inferred accretion luminosities and mass-transfer rates should be regarded strictly as lower limits.

Theoretical predictions for the amount of matter that has to be accumulated on such a WD, so that the conditions for thermonuclear runaway are met, $M_\mathrm{tnr}$, vary in the range of $(0.2\mathrm{-}3)\times10^{-6}\ M_\odot$ \citep[e.g.][]{2016ApJ...825...95D, 2020ApJ...895...70S, 2025A&A...698A.251J, 2025ApJ...982...89S}, depending on the exact parameters of the WD and the accreted material. Adopting $M_\mathrm{tnr} = 1\times10^{-6}\ M_\odot$ for simplicity, the derived lower limits on the mass transfer rate correspond to an approximately 40 year upper limit of the ``inter-outburst period'', if the mass-transfer rate remains approximately constant throughout quiescence.

Such an upper limit is comparable to the past average inter-outburst periods in the systems, which are 29 years for V3890Sgr and 39 years for V745Sco. This may indicate that stable accretion at approximately the current rate could alone be sufficient to sustain past recurrence rates, and that no period of substantially increased accretion, similar to that recently observed in T~CrB \citep[e.g.][]{2025A&A...701A.176M}, is required for the thermonuclear runaway to occur. One must, however, consider a few caveats. The derived upper limit is heavily dependent on the adopted $M_\mathrm{tnr}$. If we adopt $M_\mathrm{tnr} = 3\times10^{-6}\ M_\odot$, which is at the higher end of predicted range, the upper limit of the predicted ``inter-outburst period'' immediately increases to about 120 years. Although previously discussed scenario is not necessarily dismissed in this case, because we are discussing the upper limit, it becomes much more speculative and dependent on the amount of energy radiated shortwards of Swift/UVOT passbands. On the other hand, Eq.~\ref{eq:masstransferrate} assumes that the entirety of potential energy is transferred to accretion luminosity, which holds true if both the luminosity of the accretion disk and the boundary layer are observed. If instead we only observe the luminosity of the accretion disk, a factor of 2 is added in the numerator (as half of the gravitational potential energy is stored in the kinetic energy of the orbiting mass in the disk), halving the upper limit of the ``inter-outburst period'' to just 20 years, which further strengthens the scenario discussed above. In fact, such a short recurrence period might even suggest that the mass transfer rate at the time of the X-shooter observation is actually too high to support observed recurrence rate at a constant accretion level, and that, possibly, our observations captured one of the systems (or both) during a period of increased accretion, compared to the average of the quiescent phase. This would imply that periods of increased accretion occur throughout the quiescent phase. However, we only present this as a possible scenario, which is not preferentially supported by available data. To a lesser amount, the intervals of reasonable WD radii and masses also impact the calculated mass transfer rate, although they are to some extent already accounted for in the $M_\mathrm{tnr}$ value interval. Uncertainties on the derived $L_\mathrm{acc}{>2000\ \AA}$ for both objects must also be kept in mind when referring to the mass transfer rate lower limit. 

Such a high inferred $\dot{M}_\mathrm{acc}$ lower limit may favour Roche-lobe overflow as the dominant mass transfer mechanism if the donors are RGB stars, since their expected wind mass-loss rates are typically below the inferred accretion rate. Should the giants already be on the AGB, they could have sufficient mass-loss rates for the accretion to be powered via wind capture alone or at least predominantly. 

\begin{figure}
  \resizebox{\hsize}{!}{\includegraphics{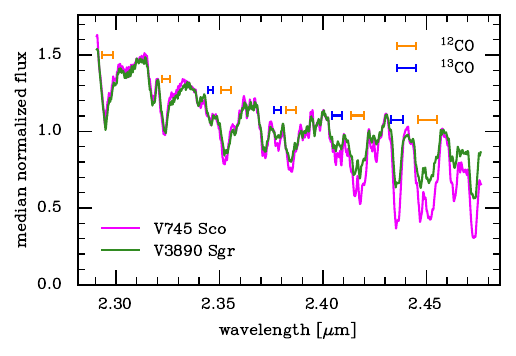}}
  \caption{Comparison of the CO band depths in the two objects used for estimating the \element[][12]{C}/\element[][13]{C} ratio. Spectrum of V745~Sco in pink, V3890~Sgr in green. Orange bars indicate positions of \element[][12]{C}\element[][]{O} bands, blue bars those of \element[][13]{C}\element[][]{O} bands. The spectra are smoothed by running mean of nearest 60 points, so that the band heads are more clearly visible.}
  \label{fig:CObands}
\end{figure}

One of the indicators of the evolutionary stage of the red giant is the \element[][12]{C}/\element[][13]{C} ratio. \citet{2022MNRAS.517.6064K} derive $\element[][12]{C}/\element[][13]{C}\sim25$ for V3890~Sgr, based on spectral fitting in the 2.25--2.50 $\mu\mathrm{m}$ range. While we do not measure \element[][12]{C}\element[][]{O} in either object directly, we find that X-shooter spectra of the two objects show very similar features in this range (see Fig.~\ref{fig:CObands}), especially considering relative depths of \element[][12]{C}\element[][]{O} and \element[][13]{C}\element[][]{O} bands. The \element[][12]{C}/\element[][13]{C} ratio in V745~Sco thus may be similar to that in V3890~Sgr. $\element[][12]{C}/\element[][13]{C}\sim25$ is consistent with the post first dredge-up theoretical prediction, and somewhat lower than the lower bound of the post third dredge-up prediction \citep[e.g.][]{2014PASA...31...30K}, which might slightly favour the giants to be on the RGB, although the diagnostic power of this comparison is limited.

\section{Radial velocities and Balmer lines}\label{sec:spectral_features}
\subsection{Radial velocity}
We determine the barycentric radial velocity of the RGs in the two systems using \texttt{iSpec} \citep{2014A&A...569A.111B, 2019MNRAS.486.2075B}. The RVs were determined from the cross-correlation function (CCF) between the observed spectrum and a synthetic template. For the latter, we use a generic synthetic RG spectrum from the spectral library of \citet{2005A&A...442.1127M}, setting the key parameters to: $T_\mathrm{eff}=3500\ \mathrm{K}$, $\log(g)=0$, $[\mathrm{M/H}] = 0.5$ and limiting the wavelength interval to 9000--10500\ $\text{\AA}$, where the RG dominates. We calculate: 
\begin{equation*}
    \mathrm{RV}^{V745~\mathrm{Sco}}_{\mathrm{RG}} = -124\pm3\ \mathrm{km\ s^{-1}}
\end{equation*}
and
\begin{equation*}
    \mathrm{RV}^{V3890~\mathrm{Sgr}}_{\mathrm{RG}} = -112 \pm3\ \mathrm{km\ s^{-1}}.
\end{equation*}
The derived RV of V3890~Sgr is consistent with the orbital solution of \citet{2021MNRAS.504.2122M}.

\subsection{Balmer lines}
The Balmer lines in both objects exhibit similar qualitative behaviour. We observe broad emission lines with a narrower absorption component, superimposed onto them. The absorption component is progressively redshifted relative to the emission as we move from Balmer~$\alpha$ to Balmer~$\delta$, resembling the behaviour observed in THA $15-31$ \citep{tha1531}. We analyse the line profiles by modelling them with a Voigt profile for the emission component and a Gaussian profile for the absorption component. Uncertainties of radial velocities reported in the rest of this section are on the order of $1\ \mathrm{km}\ \mathrm{s}^{-1}$.

\begin{figure*}
    
   \centering
   \includegraphics[width=18cm]{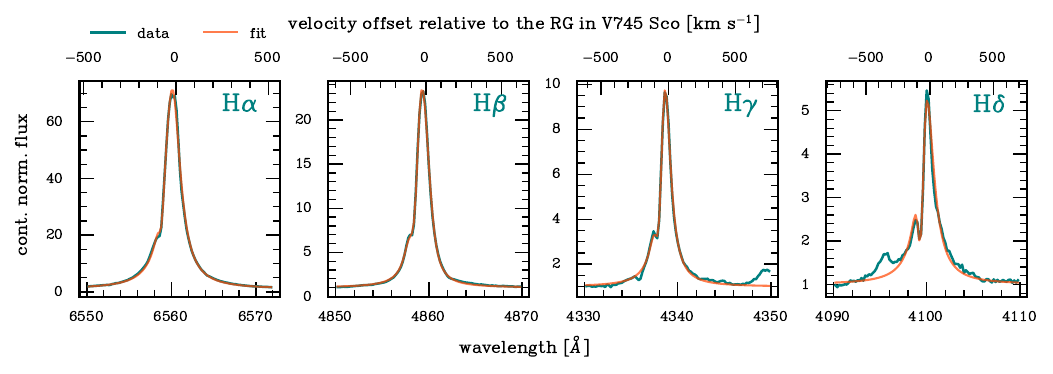}
   \caption{Continuum normalized profiles of the first four Balmer lines in the spectrum of V745 Sco. Data is presented in blue, best-fitting model in orange.}
   \label{fig:balmer_lines_V745}

\end{figure*}

The absorption component in the Balmer lines in V745~Sco is significantly blueshifted compared to the emission component, to the point that the profile could easily be interpreted as a two-peaked emission, as seen in Fig.~\ref{fig:balmer_lines_V745}. We attempt to model the profile with two emission components as well, but cannot reproduce a two-peaked profile that would also match the wings of the emission profile. The properties of the best-fitting emission-absorption model, which match observed profiles very well, are reported in Tab.~\ref{table:balmer}. We don't identify a single underlying reason that would be the cause of the progressive redshifting of the absorption component relative to the emission one. The apparent redshift of the absorption component between H$\alpha$ and H$\beta$ is caused by a blueshift of the emission component. In contrast, the shift observed between H$\beta$ and H$\gamma$ is produced by a redshift of the absorption component, while the emission component remains approximately stationary. From H$\gamma$ to H$\delta$, both the emission and absorption components remain at nearly constant velocity, and the apparent redshift of the absorption feature is instead caused by broadening of the emission line. We note that an additional emission feature at $\lambda\sim4096\ \text{\AA}$ affects the wing of H$\delta$, making the deconvolved parameters of this line less reliable.

\begin{table*}

\caption{Measured properties of Balmer lines in V745~Sco and V3890~Sgr.}              
\label{table:balmer}      
\centering                                      
\begin{tabular}{c c c c c c c c c c c c c}          
\noalign{\smallskip}
\hline
\hline
\noalign{\smallskip}
&&\multicolumn{5}{c}{V745~Sco}&&\multicolumn{5}{c}{V3890~Sgr}\\
& &\multicolumn{3}{c}{emission}&\multicolumn{2}{c}{absorption}& &\multicolumn{3}{c}{emission}&\multicolumn{2}{c}{absorption}\\\cmidrule(lr){3-5}\cmidrule(lr){6-7}\cmidrule(lr){9-11}\cmidrule(lr){12-13}
	  & & flux  &  RV & FWHM & RV & FWHM & &flux & RV &FWHM & RV & FWHM  \\
line & &erg s$^{-1}$ cm$^{-2}$ & $\mathrm{km}\ \mathrm{s}^{-1}$ & $\mathrm{km}\ \mathrm{s}^{-1}$& $\mathrm{km}\ \mathrm{s}^{-1}$& $\mathrm{km}\ \mathrm{s}^{-1}$& &erg s$^{-1}$ cm$^{-2}$& $\mathrm{km}\ \mathrm{s}^{-1}$& $\mathrm{km}\ \mathrm{s}^{-1}$& $\mathrm{km}\ \mathrm{s}^{-1}$& $\mathrm{km}\ \mathrm{s}^{-1}$\\
\cmidrule(lr){3-5}\cmidrule(lr){6-7}\cmidrule(lr){9-11}\cmidrule(lr){12-13}
H$\alpha$&&$20.3\times10^{-14}$&-20&104&-80&27&&$20.2\times10^{-13}$&56&150&32&50\\
\noalign{\smallskip} 
H$\beta$&& $2.2\times10^{-14}$ &-32&118&-81&57&&$2.6\times10^{-13}$&54&139&43&57\\
\noalign{\smallskip}
H$\gamma$&& $0.9\times10^{-14}$ &-29&106&-74&49&&$1.2\times10^{-13}$&46&107&47&65\\
\noalign{\smallskip}
H$\delta$& &$0.6\times10^{-14}$ &-28&143&-73&47&&$0.7\times10^{-13}$&54&150&53&54\\

\hline                      
\end{tabular}
\tablefoot{Reported fluxes are not dereddened, the radial velocities are reported in the frame of the red giant. Uncertainties of radial velocities are on the order of $1\ \mathrm{km}\ \mathrm{s}^{-1}$.}
\end{table*}

\begin{figure*}
   \centering
   \includegraphics[width=18cm]{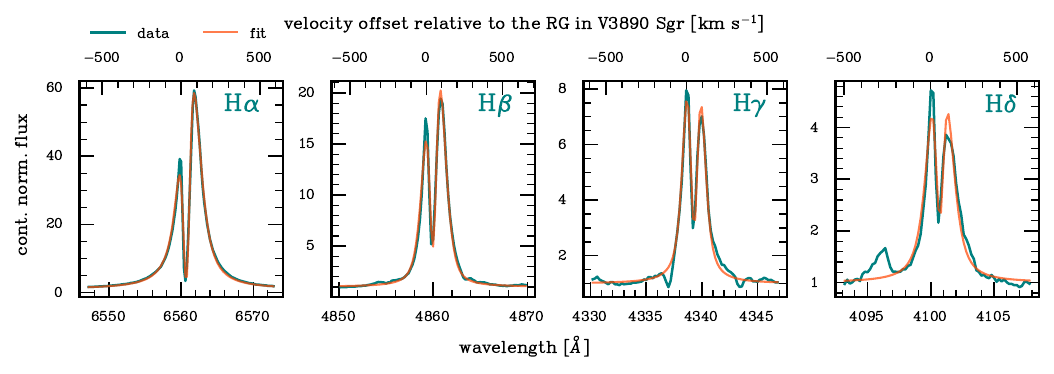}
   \caption{Continuum normalized profiles of the first four Balmer lines in the spectrum of V3890 Sgr. Data is presented in blue, best-fitting model in orange.}
   \label{fig:balmer_lines}
\end{figure*}

The Balmer lines in the spectrum of V3890~Sgr alongside the best-fitting models are presented in Fig.~\ref{fig:balmer_lines}.
In general, the best-fitting models match the emission lines' wings well, and also seem to accurately recover the central wavelength of the absorption component. They do, however, struggle to completely reproduce the depth of the absorption feature and the peaks surrounding it. Introduction of asymmetrical absorption component (a Gaussian with different standard deviations for the left and the right side) enables us to find a better fitting curve for H$\alpha$. Because the solution in that case is strongly dependent on initial parameters and limits of allowed parameter space, and because there is no clear and obvious physical justification for the asymmetry of the absorption lines, we stick with the symmetric model. We report the properties of the best-fitting models alongside those of V745~Sco in Tab.~\ref{table:balmer}. In this case, the apparent redshifting of the absorption component is indeed caused by the redshifting of the absorption component. The central wavelength of the absorption line is progressively redshifted along the Balmer series, while that of the emission component remains almost stationary in H$\alpha$, H$\beta$ and H$\delta$. Emission component in H$\gamma$ is offset by about $10\ \mathrm{km}\ \mathrm{s}^{-1}$, although both of the wings in this case are disturbed by surrounding absorption-like features, so the emission line parameters are less reliable. 

The emission line flux ratios along the Balmer series are consistent between the two objects, indicating that emission processes are similar. The change of the properties of absorption lines along the Balmer series in the two objects suggests that they are tracing different parts of spatially inhomogeneous \ion{H}{i}. Phase-resolved spectroscopy is, however, needed to properly interpret them in light of physical properties of the system.

The first five lines of the Paschen series are observed in emission in both objects, whereas no clear emission is detected in the Brackett series. Visual inspection of these lines reveals no evident behaviour that would warrant further investigation at this stage. The emission-line spectra of both objects are exceptionally rich, however, a detailed discussion of their properties lies beyond the scope of the present paper.

\section{Summary and conclusions}\label{sec:conclusion}

We investigated the quiescent states of V745~Sco and V3890~Sgr in 2024 to characterise late-type giants and the accretion properties in the systems. We find that V745~Sco and V3890~Sgr harbour M7III and M6.5III giants, respectively, classified in the Case/Nassau system. We derived extinction and distances to the giants by comparing their colours and magnitudes to those of typical bulge late-type giants. 

Using this information we constructed the SEDs of our objects and compared them to those of single bulge late-type giants to calculate properties of accretion based on UV and IR excess. We find significant UV excess in both objects on the order of $340\ L_{\odot}$, while IR excess is only present in V745~Sco, on the order of $260\ L_{\odot}$. 

Along with the radial velocities of the RGs in the systems, we also report properties of deconvolved emission and absorption features in the profiles of the lines in the Balmer series. We find that the apparent progressive redshift of the absorption component in the Balmer series spectral lines is not caused by the same physical features in both systems, tracing different regions of spatially inhomogeneous neutral hydrogen, possibly connected with different gas flows within the binaries or with different lines of sight resulting from differing orbital phases of the systems. 

We use the UV excess to derive lower limits on the mass-transfer rates in the systems to $\dot{M}_\mathrm{acc} \gtrsim 2.4\times10^{-8}\ M_\odot\ \mathrm{yr}^{-1}$ and $\dot{M}_\mathrm{acc} \gtrsim 2.5\times10^{-8}\ M_\odot\ \mathrm{yr}^{-1}$ for V745~Sco and V3890~Sgr, respectively. Such high mass transfer rates may favour Roche-lobe overflow as the dominant mass transfer mechanism if the donor is on the RGB, but do not exclude wind accretion if the giant has already reached the AGB. Assuming that $10^{-6}\ \mathrm{M}_\odot$ has to accumulate on the WD for a nova outburst to begin, we derive the upper limit on the characteristic recurrence timescale on the order of 40 years. This is comparable to the average inter-outburst periods in the systems (29 years for V3890~Sgr and 39 years for V745~Sco), suggesting that the observed recurrence times may be compatible with approximately steady accretion throughout quiescence. While this does not exclude episodes of enhanced accretion, it indicates that a prolonged enhanced-accretion phase similar to that recently observed in T~CrB is not necessarily required. In this respect, the quiescent behaviour of the two systems may be more similar to that of RS~Oph than to that of T~CrB, having no distinct well-behaved phases with different levels of accretion, but instead exhibiting similar behaviour throughout the entirety of the quiescent phase. We note that this suggestion is sensitive to several assumptions, most notably the amount of accumulated mass required for the TNR to occur and the fact that the inferred accretion rate represents only a lower limit.

We also discuss the origin of the IR excess in V745~Sco, finding that the presence of a few $10^{-9}\ M_\odot$ of dust at a characteristic temperature of $\sim780\ $K, combined with the SED of a typical bulge giant, reproduces the observed infrared excess well. On the other hand, our perfectly flat and thin disk model struggles to reproduce the observed infrared excess, but the simplicity of the considered model prevents us from dismissing the outer layers of the disk as a possible source of the IR excess. We therefore consider both explanations for the cause of the IR excess plausible. The observed post-outburst evolution in the NEOWISE data could be explained by a gradual reformation of a dusty circumstellar component in the condensing wind of the late-type giant, or with the reformation of the outer layers of the accretion disk. The reformation of the inner part of the accretion disk is suggested by the increase in the Gaia BP band 2--3 years after the outburst. It precedes the increase in NEOWISE $W1$ and $W2$ bands by about 5 years, indicating an inside-out reformation of the disk, should it be responsible for both increases.

We present the determined properties of the systems in Table~\ref{table:basic_properties}. They trace the state of the system at the epoch of our observations, and the ones pertaining to the interaction between the components in a system are expected to vary throughout the quiescent phase, as observed in V3890~Sgr before the most recent outburst \citep{2021MNRAS.504.2122M}. Continued spectrophotometric monitoring of the objects is required to better understand not only the physical properties underlying the emission features in the objects, but also overall behaviour of the quiescent phase in the systems and compare them with those of T~CrB, RS~Oph and other symbiotic stars. 

\begin{acknowledgements}
We are grateful to ANS Collaboration observers S.~Dallaporta and A.~Maitan for their support in providing the $BVRI$ photometry of V3890~Sgr. We gratefully acknowledge the work of AAVSO observer Franz-Joseph Hambsch for obtaining photometry of V745~Sco. We gratefully acknowledge the contributions of the AAVSO observer community, whose photometric data and metadata resources were used in this study and made available through the AAVSO’s scientific archives. BJ thanks Janez Kos for discussion and advice. BJ, GT and MP acknowledge the support of the Slovenian Research Agency (research core funding No. P1-0188). UM acknowledges the support by INAF 2023 MiniGrant Programme (contract C93C23008470001). NM acknowledges financial support through ASI-INAF and ’Mainstream’ agreement No. 2017-14-H.0 (PI: T. Belloni). GT acknowledges financial support from the European Space Agency (Prodex Experiment Arrangement No. 4000143450). Based on data obtained from the ESO Science Archive Facility with DOI(s): \url{https://doi.eso.org/10.18727/archive/71}. We acknowledge the use of public data from the Swift data archive. This publication makes use of data products from the Wide-field Infrared Survey Explorer, which is a joint project of the University of California, Los Angeles, and the Jet Propulsion Laboratory/California Institute of Technology, funded by the National Aeronautics and Space Administration. This publication makes use of data products from NEOWISE, which is a project of the Jet Propulsion Laboratory/California Institute of Technology, funded by the Planetary Science Division of the National Aeronautics and Space Administration. This publication makes use of data products from the Two Micron All Sky Survey, which is a joint project of the University of Massachusetts and the Infrared Processing and Analysis Center/California Institute of Technology, funded by the National Aeronautics and Space Administration and the National Science Foundation. This research was made possible through the use of the AAVSO Photometric All-Sky Survey (APASS), funded by the Robert Martin Ayers Sciences Fund and NSF AST-1412587. This work has made use of data from the European Space Agency (ESA) mission {\it Gaia} (\url{https://www.cosmos.esa.int/gaia}), processed by the {\it Gaia} Data Processing and Analysis Consortium (DPAC, \url{https://www.cosmos.esa.int/web/gaia/dpac/consortium}). Funding for the DPAC has been provided by national institutions, in particular the institutions participating in the {\it Gaia} Multilateral Agreement. This research made use of the following software: \texttt{Astropy} \citep{astropy:2013, astropy:2018, astropy:2022}, \texttt{SciPy} \citep{2020SciPy-NMeth}, \texttt{NumPy} \citep{harris2020array}, \texttt{Matplotlib} \citep{Hunter:2007}, and \texttt{pyphot} \citep{fouesneau2026pyphot}.
\end{acknowledgements}

\bibliographystyle{aa} 
\bibliography{bibliography.bib}
\begin{appendix} 
\nolinenumbers
\section{Bulge late-type giants photometry}\label{sec:app_Bulge}
\citet{1982ApJ...259L...7F} and \citet{1987ApJ...320..199F} found that bulge late-type giants appear bluer and fainter than their halo and Solar neighbourhood counterparts of the same spectral type. They conducted a photometric survey of late-type giants in the Baade window, which were reported by \citet*[BMB, M6III--M9III]{1984AJ.....89..636B} and \citet[B86, M1III--M5III]{1986AJ.....91..290B}, and classified based on the Case system. A key result of the \citet{1987ApJ...320..199F} paper is a set of tables of intrinsic colours (and luminosities) of late-type bulge giants. Unfortunately, the $JHK$ passbands used in their survey are significantly different than those used by 2MASS \citep{2003yCat.2246....0C}, and a direct comparison to the 2MASS colours of our two systems would introduce an unknown systematic offset. To mitigate such effects and expand the table to the WISE passbands, we decide to construct similar tables ourselves. 

Using the classification of giants by BMB and B86, we collect $V$ and $I$ band photometry from OGLE-II (\citealt{1997AcA....47..319U,2002AcA....52..217U,2005AcA....55...43S}, object identification from \citealt{2011A&A...529A.104C}) and OGLE-III photometric maps of the bulge \citep{2008AcA....58...69U,2013AcA....63...21S}, $J$, $H$ and $K_{\rm s}$ from 2MASS, and $W1$, $W2$, $W3$ and $W4$ from AllWISE catalogue \citep{2010AJ....140.1868W} for all of the giants within a given spectral type. Unexpectedly, in addition to the spectral types of giants reported in FW87, we also find that a number of objects are classified as M6.5III giants, while no other half-classes are present. Comparing the reported number of objects per spectral type in FW87 (their Table~2) to our Table~\ref{table:number_of_systems}, we suspect that most BMB objects classified as M6.5III were considered M6III objects in the FW87 analysis.

We deredden the colours using prescriptions from \citet{2003A&A...401..781F} for OGLE and 2MASS passbands, and prescriptions from \citet{2023ApJS..264...14Z} for WISE bands, taking $E(B-V) =0.48$ for all of the stars in the Baade window, as done by FW87. Testing for the effect of differential extinction using \citet{2011ApJ...737..103S} recalibration of \citet{1998ApJ...500..525S} (SFD) map of total galactic extinction, we find differences on the order of 0.03 magnitude in reported mean colours, with no consistent effect on the observed scatter. We stick with uniform reddening for consistency with FW87 and because SFD map reports total galactic extinction, which may exceed extinction at the $\sim8\ \mathrm{kpc}$ distance to the bulge.  

We filter out magnitudes flagged as upper limit detections in AllWISE and 2MASS data. In the latter, we also filter out detections flagged for very poor goodness of PSF fit. OGLE-II $I-$band magnitudes are systematically offset for very red objects, as reported in \citet{2011AcA....61...83S}. We apply their recalibration to account for this. Reported formal uncertainties on 2MASS and AllWISE data are on the order of 0.02 magnitude, while OGLE reports amplitude of the variability of the magnitudes, which range from a couple hundredths of magnitude to approximately half a magnitude. The observed scatter in the colour distributions is thus most likely predominantly caused by intrinsic scatter of the giants' colours and their variability (which is important due to non-simultaneous observations), rather than uncertainties of individual observations or assumptions regarding the extinction to the Baade window.

\begin{figure*}

   \centering

   \includegraphics[width=18cm]{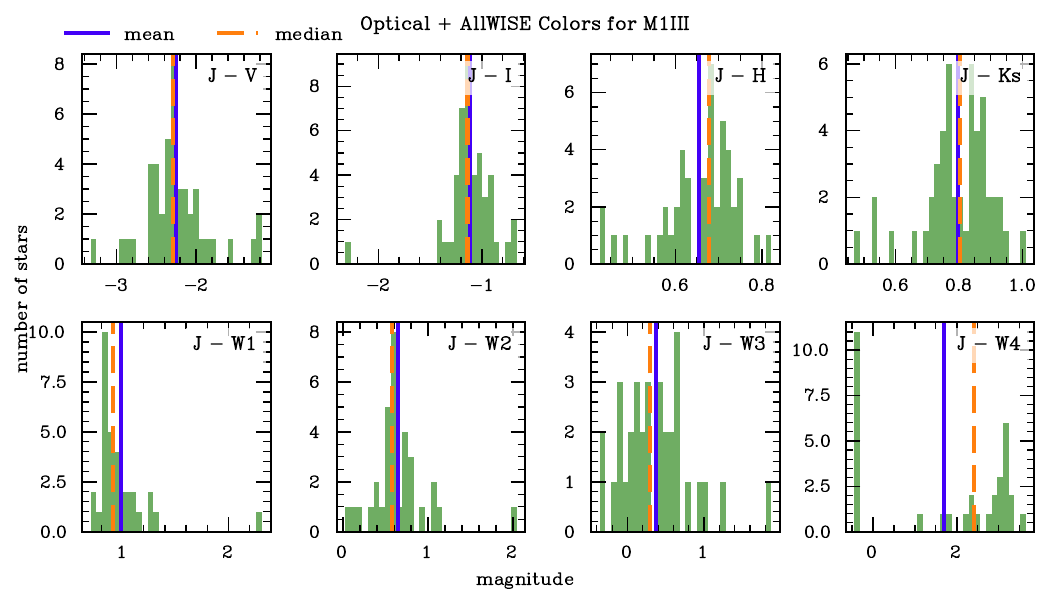}
   \caption{Distributions of dereddened colours of M1III type giants in the Baade window. Blue vertical solid lines mark mean colours in the sample, orange dashed lines mark median colours in the sample.}
   \label{fig:M1III_bulge_hist}

\end{figure*}

We present intrinsic mean and median colours of bulge late-type giants in Tables~\ref{tab:bulge_giants_mean}~and~\ref{tab:bulge_giants_median}, respectively. Along the mean colours in Table~\ref{tab:bulge_giants_mean} we also report the standard deviation of colours within the sample. We supply plots of distributions of all colours for each spectral type in Figs.~\ref{fig:M1III_bulge_hist}~to~\ref{fig:M9III_bulge_hist}, so that the readers may themselves decide whether mean or median colours are appropriate for their use case. 

\FloatBarrier

\begin{table*}
\caption{Number of bulge giants of each spectral type in Baade window, reported by BMB and B86, for which we collected data in OGLE, 2MASS, and AllWISE databases.} 
\label{table:number_of_systems} 
\centering 
\begin{tabular}{c *{10}{c}} 
\noalign{\smallskip}
\hline
\hline
\noalign{\smallskip}
Sp. Type & M1III & M2III & M3III & M4III & M5III & M6III & M6.5III & M7III & M8III & M9III \\
\noalign{\smallskip}
\hline
\noalign{\smallskip}
N & 52 & 35 & 23 & 16 & 22 & 112 & 98 & 54 & 11 & 5 \\
\noalign{\smallskip}
\hline 
\end{tabular}
\end{table*}

\begin{table*}
 \caption[]{Mean intrinsic colours (and the standard deviations) of bulge M-giants in the Baade window, built from OGLE, 2MASS, and AllWISE data.}
 \label{tab:bulge_giants_mean}
 \begin{center}
\begin{tabular}{cccccccccc}
\hline
\hline
\noalign{\smallskip}
Sp. Type & $J-V$ & $J-I$  & $J-H$ & $J-K_\mathrm{s}$ & $J-W1$ & $J-W2$ & $J-W3$ & $J-W4$ \\
\noalign{\smallskip}
\hline
\noalign{\smallskip}
M1III & $-2.34 \pm 0.54$ & $-1.19 \pm 0.33$ & $0.66 \pm 0.09$ & $0.80 \pm 0.11$ & $1.06 \pm 0.42$ & $0.71 \pm 0.47$ & $0.57 \pm 0.58$ & $1.40 \pm 0.26$ \\
M2III & $-2.50 \pm 0.37$ & $-1.36 \pm 0.86$ & $0.70 \pm 0.06$ & $0.86 \pm 0.07$ & $1.16 \pm 0.34$ & $0.79 \pm 0.39$ & $0.84 \pm 0.62$ & $3.78 \pm 0.58$ \\
M3III & $-3.08 \pm 0.38$ & $-1.45 \pm 0.41$ & $0.77 \pm 0.04$ & $0.97 \pm 0.06$ & $1.11 \pm 0.11$ & $0.83 \pm 0.15$ & $0.89 \pm 0.38$ & $2.81 \pm 0.35$ \\
M4III & $-3.51 \pm 0.63$ & $-1.67 \pm 0.64$ & $0.77 \pm 0.04$ & $1.00 \pm 0.08$ & $1.14 \pm 0.14$ & $0.84 \pm 0.17$ & $0.96 \pm 0.39$ & $2.83 \pm 1.22$ \\
M5III & $-3.96 \pm 0.42$ & $-1.71 \pm 0.14$ & $0.81 \pm 0.04$ & $1.08 \pm 0.05$ & $1.19 \pm 0.08$ & $0.97 \pm 0.12$ & $1.22 \pm 0.30$ & $2.17 \pm 1.02$ \\
M6III & $-4.89 \pm 0.72$ & $-2.07 \pm 0.41$ & $0.85 \pm 0.07$ & $1.14 \pm 0.08$ & $1.30 \pm 0.22$ & $1.08 \pm 0.24$ & $1.40 \pm 0.39$ & $2.16 \pm 0.79$ \\
M6.5III & $-5.79 \pm 0.89$ & $-2.46 \pm 0.37$ & $0.87 \pm 0.09$ & $1.20 \pm 0.14$ & $1.43 \pm 0.36$ & $1.22 \pm 0.42$ & $1.66 \pm 0.58$ & $2.28 \pm 0.85$ \\
M7III & $-6.58 \pm 1.03$ & $-2.88 \pm 0.51$ & $0.90 \pm 0.06$ & $1.28 \pm 0.11$ & $1.46 \pm 0.21$ & $1.29 \pm 0.25$ & $1.99 \pm 0.62$ & $2.69 \pm 0.87$ \\
M8III & $-7.75 \pm 0.59$ & $-3.44 \pm 0.25$ & $0.96 \pm 0.05$ & $1.40 \pm 0.07$ & $1.61 \pm 0.21$ & $1.41 \pm 0.15$ & $2.47 \pm 0.28$ & $3.16 \pm 0.44$ \\
M9III & $-7.70 \pm 0.48$ & $-3.52 \pm 0.26$ & $0.93 \pm 0.04$ & $1.38 \pm 0.07$ & $1.80 \pm 0.41$ & $1.56 \pm 0.24$ & $2.61 \pm 0.49$ & $3.36 \pm 0.76$ \\
\hline
\end{tabular}

 \end{center}
 \end{table*}

 \begin{table*}
 \caption[]{Median intrinsic colours of bulge M-giants in the Baade window, built from OGLE, 2MASS, and AllWISE data.}
 \label{tab:bulge_giants_median}
 \begin{center}
\begin{tabular}{cccccccccc}
\hline
\hline
\noalign{\smallskip}
Sp. Type & $J-V$ & $J-I$  & $J-H$ & $J-K_\mathrm{s}$ & $J-W1$ & $J-W2$ & $J-W3$ & $J-W4$ \\
\noalign{\smallskip}
\hline
\noalign{\smallskip}
M1III & $-2.31$ & $-1.16$ & $0.68$ & $0.83$ & $0.93$ & $0.58$ & $0.41$ & $1.40$ \\
M2III & $-2.53$ & $-1.21$ & $0.70$ & $0.86$ & $1.10$ & $0.73$ & $0.75$ & $4.18$ \\
M3III & $-2.98$ & $-1.37$ & $0.77$ & $0.98$ & $1.10$ & $0.85$ & $0.96$ & $2.84$ \\
M4III & $-3.39$ & $-1.40$ & $0.76$ & $1.02$ & $1.14$ & $0.85$ & $0.93$ & $2.63$ \\
M5III & $-3.91$ & $-1.69$ & $0.82$ & $1.08$ & $1.17$ & $0.94$ & $1.16$ & $1.82$ \\
M6III & $-4.82$ & $-1.99$ & $0.84$ & $1.13$ & $1.25$ & $1.01$ & $1.31$ & $2.11$ \\
M6.5III & $-5.87$ & $-2.46$ & $0.87$ & $1.20$ & $1.32$ & $1.12$ & $1.49$ & $2.13$ \\
M7III & $-6.60$ & $-2.91$ & $0.89$ & $1.25$ & $1.38$ & $1.24$ & $1.83$ & $2.67$ \\
M8III & $-7.96$ & $-3.53$ & $0.96$ & $1.42$ & $1.58$ & $1.40$ & $2.49$ & $3.19$ \\
M9III & $-7.76$ & $-3.57$ & $0.94$ & $1.42$ & $1.56$ & $1.49$ & $2.74$ & $3.38$ \\
\hline
\end{tabular}

 \end{center}
 \end{table*}
 \begin{figure*}
    
   \centering

   \includegraphics[width=18cm]{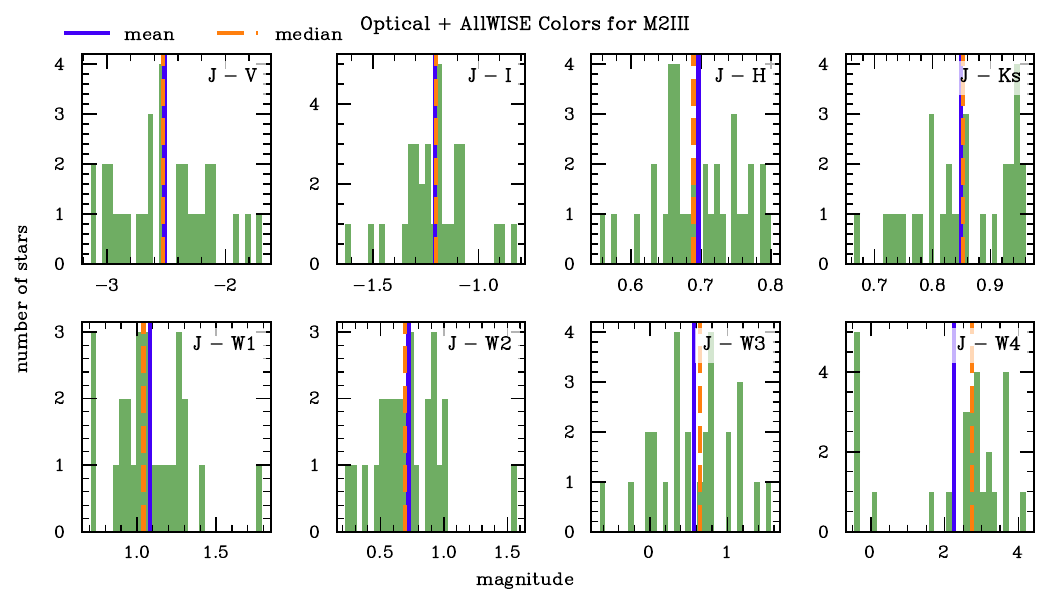}
   \caption{Same as Fig.~\ref{fig:M1III_bulge_hist}, but for M2III.}

\end{figure*}
 \begin{figure*}
    
   \centering

   \includegraphics[width=18cm]{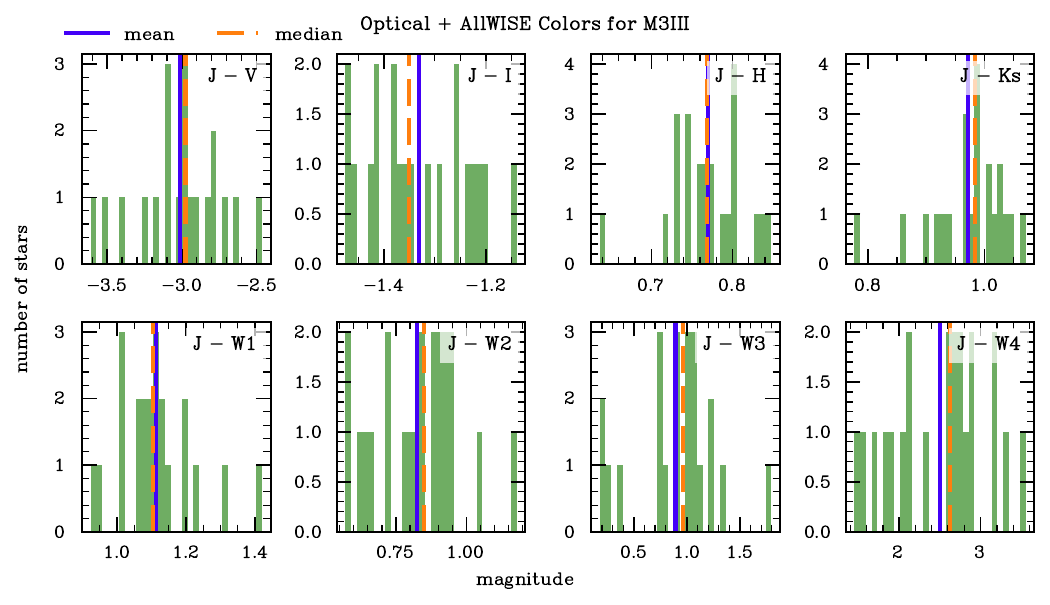}
   \caption{Same as Fig.~\ref{fig:M1III_bulge_hist}, but for M3III.}

\end{figure*}
 \begin{figure*}
    
   \centering

   \includegraphics[width=18cm]{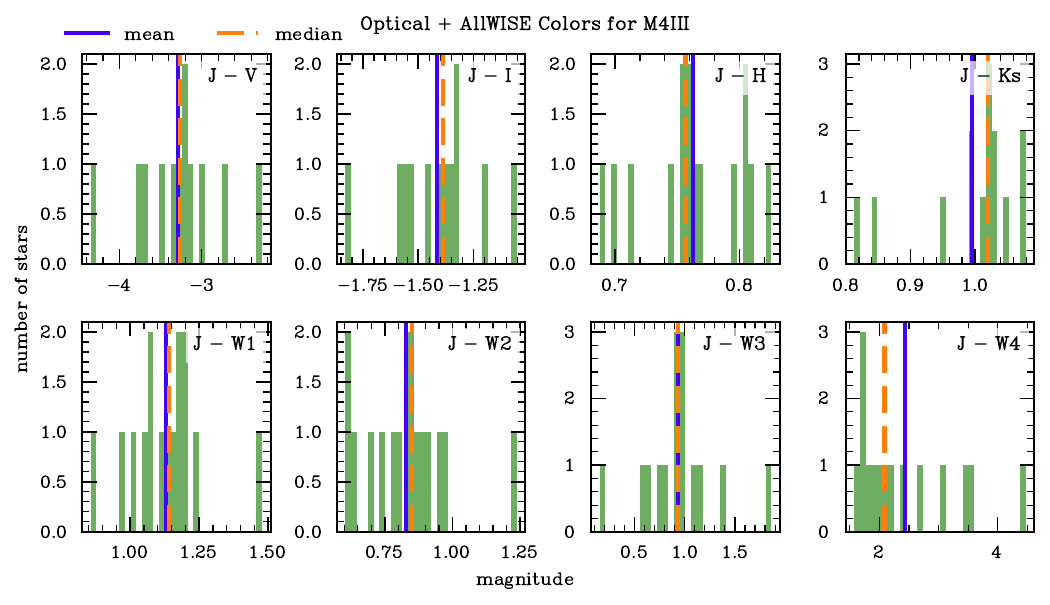}
   \caption{Same as Fig.~\ref{fig:M1III_bulge_hist}, but for M4III.}

\end{figure*}
\begin{figure*}
    
   \centering
 
   \includegraphics[width=18cm]{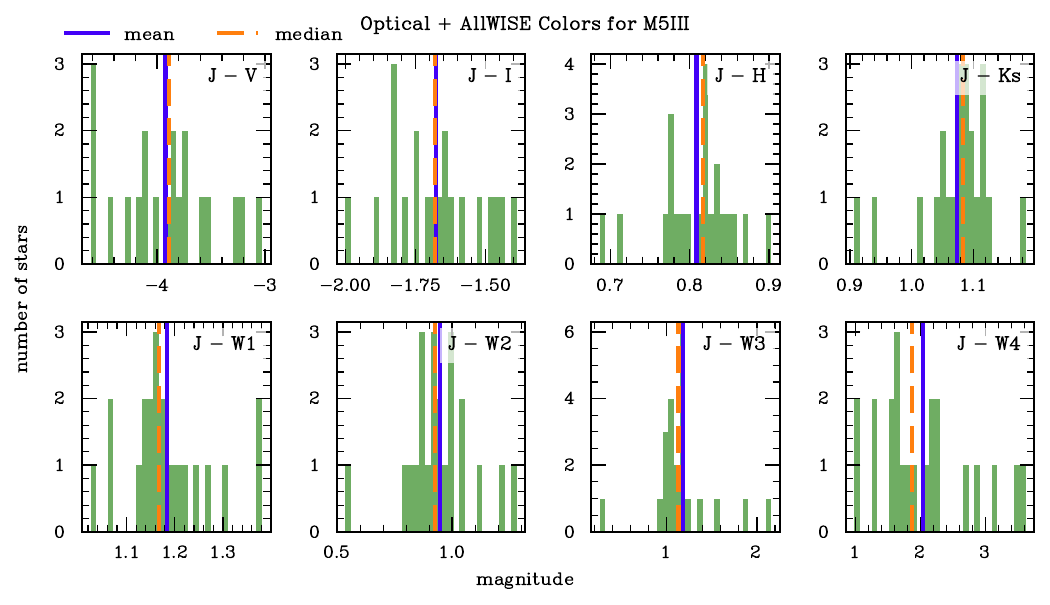}
   \caption{Same as Fig.~\ref{fig:M1III_bulge_hist}, but for M5III.}

\end{figure*} 
\begin{figure*}
    
   \centering

   \includegraphics[width=18cm]{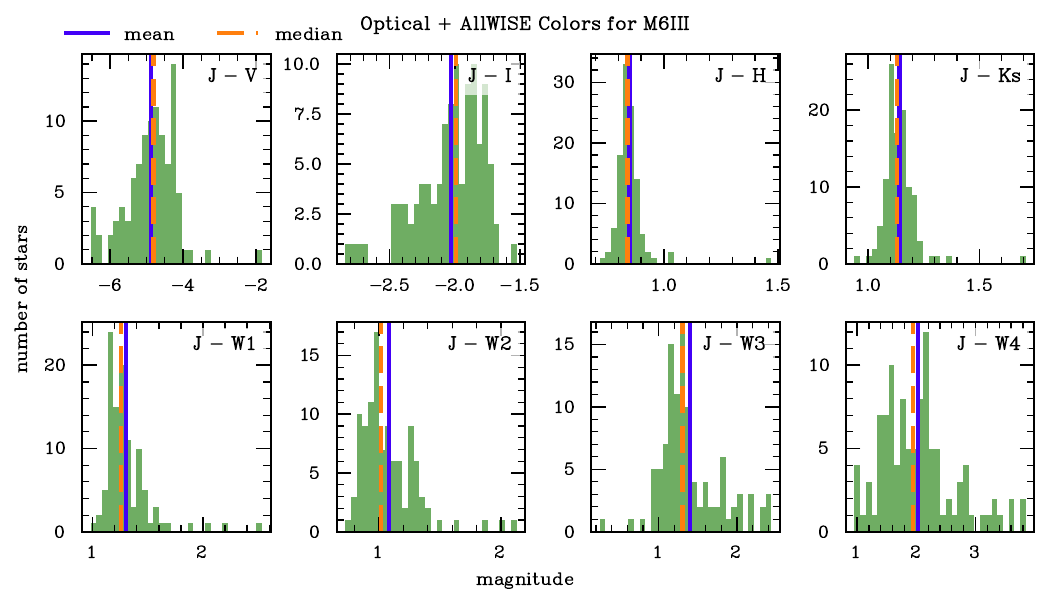}
   \caption{Same as Fig.~\ref{fig:M1III_bulge_hist}, but for M6III.}

\end{figure*}
\begin{figure*}
    
   \centering

   \includegraphics[width=18cm]{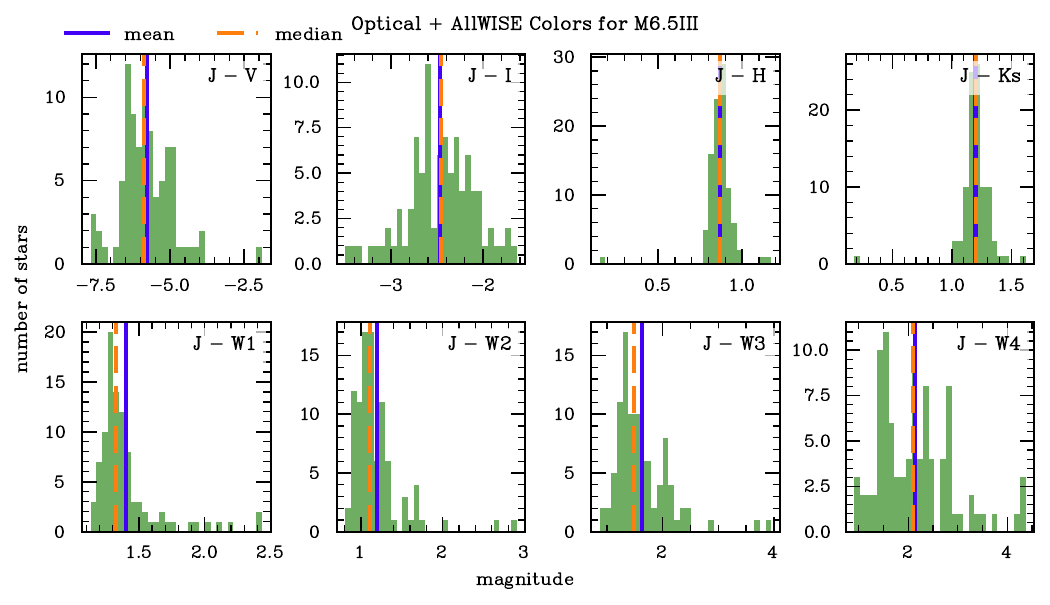}
   \caption{Same as Fig.~\ref{fig:M1III_bulge_hist}, but for M6.5III.}

\end{figure*}
\begin{figure*}
    
   \centering

   \includegraphics[width=18cm]{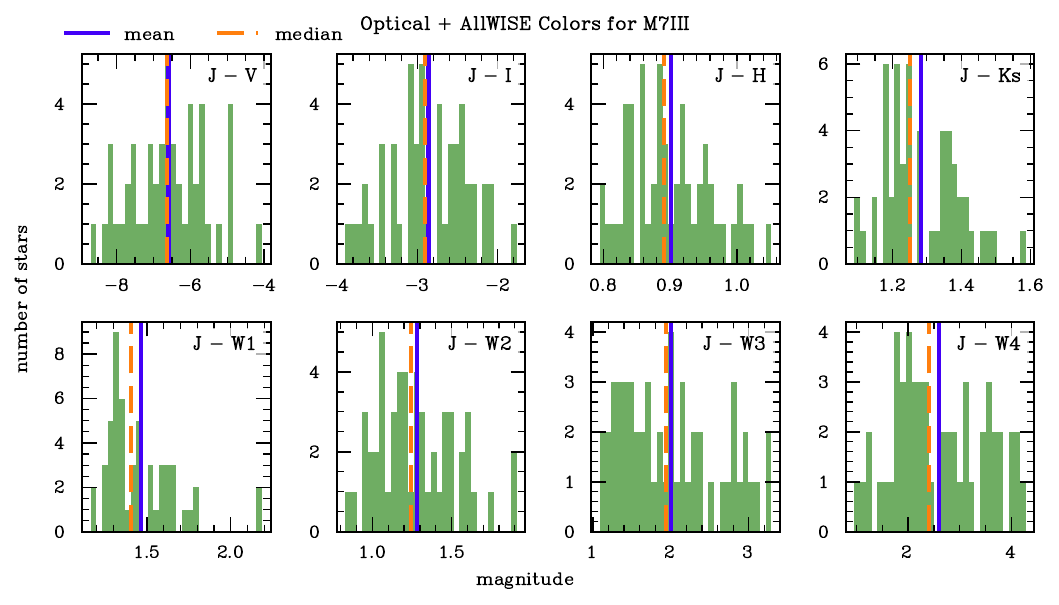}
   \caption{Same as Fig.~\ref{fig:M1III_bulge_hist}, but for M7III.}

\end{figure*}
\begin{figure*}
    
   \centering

   \includegraphics[width=18cm]{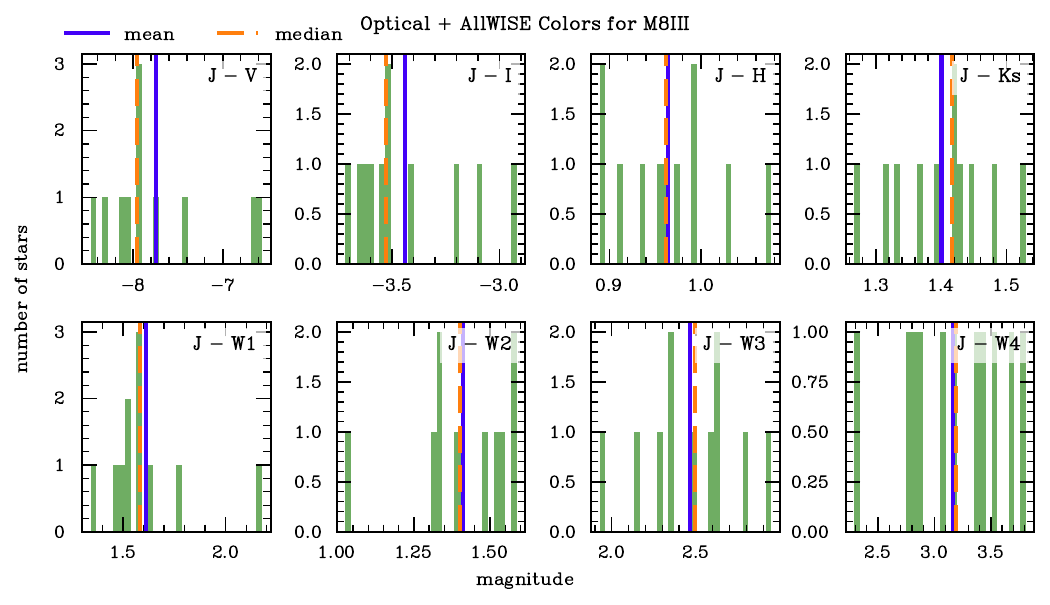}
   \caption{Same as Fig.~\ref{fig:M1III_bulge_hist}, but for M8III.}

\end{figure*}
\begin{figure*}
    
   \centering

   \includegraphics[width=18cm]{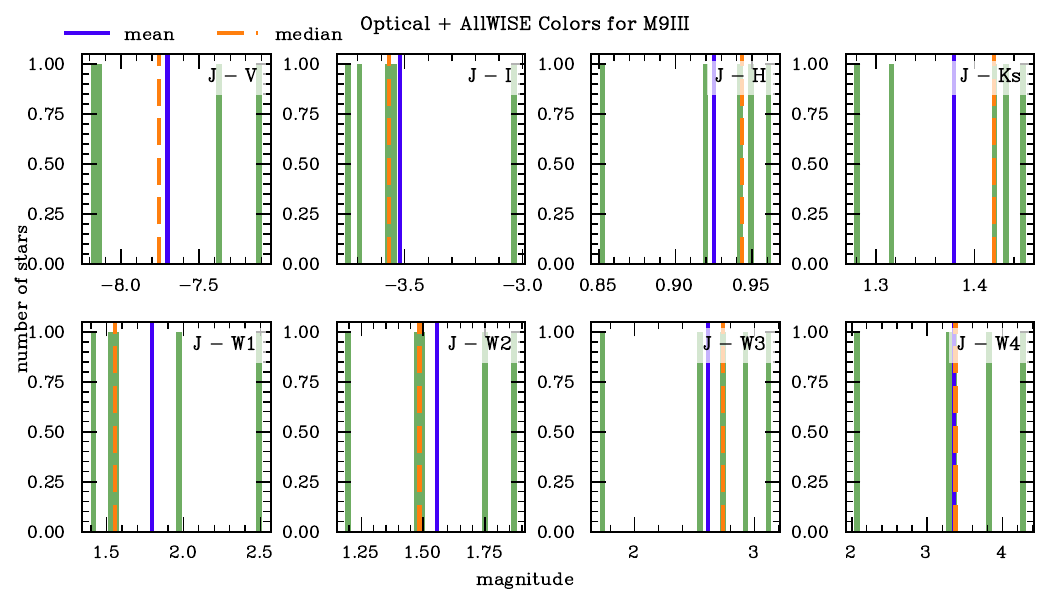}
   \caption{Same as Fig.~\ref{fig:M1III_bulge_hist}, but for M9III.}
   \label{fig:M9III_bulge_hist}

\end{figure*}
\FloatBarrier
\clearpage
\section{Systematic uncertainties}\label{app:systematic_uncertainties}
At least two sources of systematic uncertainties are present throughout the paper. The first one is stemming from spectral type misclassification (e.g. if we mistakenly classify the RG in the system as M7III, when it is actually of M6.5III spectral type). The other stems from our assumption that the RGs in the system resemble those in the bulge, when, in reality, they resemble RGs in the Solar neighbourhood.
\subsection{Effect on interstellar extinction estimate}
We estimate the contribution of spectral misclassification to our $E(B-V)$ estimates by comparing derived mean $J-K_\mathrm{s}$ colour of M7III giant to that of adjoining spectral types and find the systematic uncertainty to be of the order of $0.08$ magnitude (see Table~\ref{tab:bulge_giants_mean}). It is somewhat less straightforward to estimate the contribution stemming from the potential misidentification of the systems as belonging to the bulge population. In Fig.~10 of \citet{1987ApJ...320..199F} a difference of about $0.15$ magnitude is visible in $J-K$ giant colours when comparing bulge to non-bulge stars. However, the photometric system is not identical to that of 2MASS, which could lead to some additional offset. We also compare derived bulge giant 2MASS colours to those of field giants listed in \citet{2009BaltA..18...19S}. The comparison is again imperfect, as they do not report colours for spectral types later than M5III, and they use MK classification system, which can not be reliably translated to Case/Nassau. That being said, the difference is again on the order of $0.15$ magnitude, which we consider to be a fair estimate of the contribution to uncertainty. Propagating to $E(B-V)$, we find a systematic offset of $^{+0.16}_{-0.14}$ magnitude, corresponding to spectral misclassification and a potential $-0.26$ magnitude offset, corresponding to the possibility that the giant does not resemble bulge late-type giants. For V745~Sco we therefore calculate $E(B-V)=0.87\pm0.2_\mathrm{stat}\,^{+0.16}_{-0.14\,\mathrm{SpT}}\,-0.26$\,(bulge/non-bulge offset), differentiating between statistical uncertainty and both types of systematic offsets.
For V3890~Sgr we use $E(B-V)$ value which is directly measured from the EWs of the \ion{Na}{i}~D doublet lines, so the systematic uncertainties of the same type are not applicable. If we were to use the $E(B-V)$ which was derived in the same way as described for V745~Sco, the systematic estimates would be the same.
\subsection{Propagation to other results}
Propagating these systematic uncertainties to other results is not quite as trivial as simply summing up the uncertainties in quadrature, and just sampling from wider distributions, as these uncertainties are not independent of other assumptions. When calculating distance, a systematic offset on extinction, stemming from misclassification to a neighbouring spectral type, is always accompanied by a systematic offset on typical apparent magnitude: a typical giant of later spectral type will, for example, appear both redder and brighter than an earlier one. Thus, the comparison-derived $E(B-V)$ will be lower, which would in turn then cause the derived distance to be larger, while the fact that the comparison object is intrinsically brighter would magnify the effect even further.
Attempting to robustly determine systematic uncertainties of all of our results would be overly complex and would necessitate, for example, modelling the correlation between systematic uncertainty of model $J-K_\mathrm{s}$ colour to the overall shape of the model SED. To be able to understand the magnitude of this systematic effect, we instead again calculate the derived properties of V745~Sco, this time assuming that the red giant in the system is M6.5III type instead of M7III type, and accordingly substituting typical $J-K_\mathrm{s}$, $K_{s}$ and model SED. We compare the results in Table~\ref{tab:systematic_uncertainties}. 
\begin{table}
\caption{Comparison of properties of V745~Sco, depending on the assumed giant spectral type.}              
\label{tab:systematic_uncertainties}      
\centering                                      
\begin{tabular}{c c c c}          
\noalign{\smallskip}
\hline
\hline
\noalign{\smallskip}
RG Sp. Type&M7III&M6.5III&unit\\
\noalign{\smallskip}
\hline
\noalign{\smallskip} 
$E(B-V)$   & $0.87\pm0.20$ & $1.01\pm0.25$ &mag\\
\noalign{\smallskip}
Distance  & $8.5\,^{+3.1}_{-2.3}$ & $6.7\,^{+2.6}_{-1.9}$&kpc\\
\noalign{\smallskip}
$L_\mathrm{UVOT}$ & $56^{+197}_{-43}$&$93^{+494}_{-78}$& $L_{\odot}$\\
\noalign{\smallskip}
$L^{>2000\,\AA}_\mathrm{acc}$ & $340^{+690}_{-240}$ &$360^{+1210}_{-290}$ & $L_{\odot}$\\
\noalign{\smallskip}
$\dot{M}_\mathrm{acc}$ &  $\gtrsim2.4\times10^{-8}$ & $\gtrsim2.5\times10^{-8}$ & $M_{\odot}\,\mathrm{yr}^{-1}$\\
\noalign{\smallskip}
$L_\mathrm{IR}$ & $260^{+270}_{-140}$&$150^{+190}_{-90}$  & $L_{\odot}$\\
\noalign{\smallskip}
\hline                      
\end{tabular}
\end{table}
As expected, derived $E(B-V)$ and distance are different, although they agree within their uncertainties. Interestingly, however, other properties are less affected, their expected values remaining closer together. However, their uncertainties do grow, which is likely due to somewhat larger scatter on model $J-K_\mathrm{s}$ and $K_{s}$. The differences do not impact our results significantly enough to change our conclusions qualitatively. 
\end{appendix}
\end{document}